\documentclass[aps,prd,preprint,showpacs,amsmath,amssymb,amsfons]{revtex4-1}
\usepackage{booktabs}

\usepackage{mathrsfs}
\usepackage{epsfig}
\usepackage{graphicx}     
\usepackage{dcolumn}      
\usepackage{bm}           
\usepackage{slashed}       
\usepackage{multirow}
\usepackage[
			colorlinks=true,
            linkcolor=blue,
            urlcolor=red,
            citecolor=blue]{hyperref}
\usepackage{color}
\usepackage{caption}
\usepackage{subcaption}
\usepackage{ulem}
\newcommand{\add}[1]{#1}

\begin{document}
\title{Single production of vector-like top partner decaying to \texorpdfstring{$Wb$}{Lg} in the leptonic channel at \texorpdfstring{$ep$}{Lg} colliders in the LHT model}
\author{Bingfang Yang$^{1}$}\email{yangbingfang@htu.edu.cn}
\author{Biaofeng Hou$^{1}$}\email{resonhou@zknu.edu.cn}
\author{Huaying Zhang$^{1}$}
\author{Ning Liu$^{2}$}
\affiliation{$^1$College of Physics and Materials Science, Henan Normal
	University, Xinxiang 453007, China\\
	$^2$Department of Physics and Institute of Theoretical Physics,
	Nanjing Normal University, Nanjing 210023, China
	\vspace*{1.5cm}  }
\date{\today}

\begin{abstract}
	
    In the littlest Higgs model with T-parity(LHT), we study the single production of vector-like top partner with the subsequent decay $T_{+}\to Wb$ in the leptonic channel at the $ep$ colliders. Focus on the LHeC ($ \sqrt{s} $ = 1.98 TeV) and FCC-eh ($ \sqrt{s} $ = 5.29 TeV), we investigate the observability of the single top partner production with the unpolarized and polarized electron beams, respectively. As a result, the statistical significance can be enhanced by the polarized electron beams.
    Under the current constraints, the search for $T_{+}$ in the $Wb$ channel at the LHeC cannot provide a stronger limit on the top partner mass. By contrast, the search for the $T_{+}$ in this channel at the FCC-eh with polarized $ e^- $ beams can exclude the top partner mass up to 1350 GeV, 1500 GeV and 1565 GeV with integrated luminosities of 100 fb$^{-1}$, 1000 fb$^{-1}$ and 3000 fb$^{-1}$ at the 2$\sigma$ level, which is an improvement with respect to the current indirect searches and the LHC direct searches. \add{Furthermore, we also give an extrapolation to the high-luminosity LHC with $\sqrt{s}=14$ TeV and $ L=3000~\rm{fb}^{-1} $. Our results
    show that the LHT model is still a natural solution to the shortcomings of the electroweak
    and scalar sector although it has been constrained severely.}
	
\end{abstract}

\pacs{14.65.Jk, 12.60.-i, 13.60.-r, 13.88.+e}

\maketitle

\tableofcontents    

\newpage
\section{Introduction}
The standard model(SM) of particle physics, in particular after the discovery of the SM-like Higgs boson in 2012 \cite{ATLAS:2012,CMS:2012}, has achieved great success. However, the SM cannot describe all phenomena observed so far and still has some theoretical problems in itself. The most notable one is the hierarchy problem caused by the Higgs mass quadratic divergence \cite{naturalness}, which has attracted a lot of attentions and has been the main
guideline for possible model building of new physics beyond the SM. Among these models, the littlest Higgs model with T-parity(LHT) \cite{Cheng:2003ju,Cheng:2004yc,Low:2004xc} is a powerful candidate.

The LHT model constructs the Higgs boson as a pseudo-Nambu-Goldstone boson and introduces the $T$-parity to avoid the constraint from the electroweak precision observables(EWPO). In this model, the quadratic divergence contributions to Higgs mass from the SM particles are cancelled by the corresponding heavy partners. Here, an additional vector-like top partner ($ T_+ $) with $T$-even quantum number is introduced to cancel the largest contribution induced by the top quark loop. Except for the different spins, this fermionic top partner has different parities from the supersymmetric (SUSY) scalar top partner (stop) \cite{supersymmetry}, i.e.,  the former is $T$-even and the latter is $R$-odd. So, we expect the SUSY stop searches at the LHC have little affect on this $T$-even top partner searches. So far, many searches for the vector-like top partner at the LHC have been performed by ATLAS \cite{VLQ-ATLAS} and CMS \cite{VLQ-CMS}, and no excess above the SM expectation is observed. As a consequence, they give the strongest limits on the top partners. Meanwhile, the relevant phenomenological studies have been performed widely\cite{VLQ-phe}. \add{In the future, the LHC will be upgraded to the High Luminosity LHC (HL-LHC) with $ \sqrt{s}=14 $ TeV and the integrated luminosity $ L=3000~\rm{fb}^{-1} $ \cite{Atlas:2019qfx}. In addition, there are other collider schemes being proposed to search for new physics, such as high energy hadron colliders: 
High Energy LHC(HE-LHC)\cite{Atlas:2019qfx}, 
Future Circular Hadron Collider(FCC-hh)\cite{FCC} and 
Super proton-proton Collider(SppC)\cite{CEPC},
as well as lepton colliders:
International Linear Collider(ILC)\cite{ILC}, 
Compact Linear Collider(CLIC)\cite{CLIC} and 
Circular Electron Positron Collider(CEPC)\cite{CEPC}. 
At these colliders, the larger events and higher accuracy will be achieved, which will provide a good opportunity for measuring the observables precisely and probing the new physics effects.}

At the LHC, the dominant production modes of the top partners are pair-produced and they usually suffer from the SM top quark backgrounds. On the other hand, in view of the great achievements at the Hadron-Electron Ring Accelerator (HERA) \cite{HERA}, the future high-energy $ep$ colliders will give us a whole new scene and has drawn wide attentions \cite{epwork}. The related physics is concerned with new phenomena possibly occurring in the fusion of electrons and partons at TeV energies. These colliders can provide higher collision energies than the $e^{+}e^{-}$ colliders and cleaner environment than the $pp$ colliders. At present, the proposed $ep$ collider is the Large Hadron Electron Collider(LHeC) \cite{LHeC}, which is designed to collide a $60\sim 140$ GeV electron beam with a 7 TeV proton beam from the LHC. This may later be extended to Future Circular electron-hadron Collider (FCC-eh) \cite{FCC-eh}, which features a 50 TeV proton beam from the FCC-hh. Furthermore, the electron beam can be polarized and has an enormous scope to probe electroweak and Higgs physics. At this kind of colliders, the dominant production modes of the top partner will be singly produced. In this paper, we will study the observability of the single top partner production at the $ep$ colliders in the LHT model.

The structure of the paper is as follows: In Sec.II, we briefly review the top partner in the LHT model. In Sec.III, we study the single production of top partner $ T_+ $ followed by $ Wb $ in the leptonic channel at the $ep$ colliders including the LHeC and FCC-eh. Finally, we give a short summary in Sec.IV.

\section{Top Partner in The LHT Model}

In this section, we will briefly review the LHT model, the details of the model and phenomenology can be found in Ref.\cite{LHTdetails}. This model is based on a $ SU(5)/SO(5) $ non-linear sigma model. At scale $ f \sim \mathcal{O} $(TeV), the global symmetry $ SU(5) $ is broken down to $ SO(5) $ by the vacuum expectation value(VEV) of the $\Sigma $ field:
\begin{align}
	\label{vev}
	\Sigma_0=\begin{pmatrix} & & {\bf 1}_{2\times 2}\\ &1&\\  {\bf 1}_{2\times 2}&& \end{pmatrix} 
\end{align}
The gauge group is assigned to be the subgroup of $ SU(5) $ as $ G_1 \times G_2 = [SU(2) \times U(1)]_1 \times [SU(2) \times U(1)]_2 $, which is broken down to the diagonal SM electroweak gauge group $ SU(2)_L \times U(1)_Y $ by the coincident VEV in Eq.(\ref{vev}). After the symmetry breaking, there arise 4 new heavy gauge bosons $W_{H}^{\pm},Z_{H},A_{H}$ whose masses are given at $\mathcal O(v^{2}/f^{2})$ by
\begin {align}
M_{W_{H}}=M_{Z_{H}}=gf(1-\frac{v^{2}}{8f^{2}}),~~M_{A_{H}}=\frac{g'f}{\sqrt{5}}
(1-\frac{5v^{2}}{8f^{2}})
\end {align}
with $g$ and $g'$ being the SM $SU(2)_L$ and $U(1)_Y$ gauge couplings, respectively. The lightest $T$-odd particle $A_{H}$ can serve as a candidate for dark matter (DM). Note that the VEV $v$ needs to be redefined as:
\begin{align}
v = \frac{f}{\sqrt{2}} \arccos{\left( 1 -
	\frac{v_\textrm{SM}^2}{f^2} \right)} \simeq v_\textrm{SM} \left( 1 +
\frac{1}{12} \frac{v_\textrm{SM}^2}{f^2} \right)\,
\end{align}
where $v_\textrm{SM} = 246 \textrm{GeV}$ is the SM Higgs VEV. 

In order to avoid the strong constraints from the EWPO, a feasible way is to impose a discrete symmetry called $T$-parity
in this model, which plays a similar role as $R$-parity in SUSY. 
Apart from the scalar and gauge sectors, the $T$-parity also has to be implemented in the fermion sector so that every SM fermion has a mirror partner with $T$-odd quantum number. 

In the top quark sector, an additional vector-like $T$-even top partner $ T_+ $ is introduced for cancelling the large one-loop quadratic divergences of Higgs mass caused by the top quark. Then, the implementation of $T$-parity also requires its own $T$-odd mirror partner $ T_- $. The top quark has to be represented as an incomplete $ SU(5) $ multiplet $ Q_1 $ and its $T$-parity partner $ Q_2 $:
\begin{align}
	Q_1 &= \begin{pmatrix} \psi_1 \\ t'_{1} \\ {\bf 0} \end{pmatrix} \qquad \qquad
	Q_2 = \begin{pmatrix} {\bf 0} \\ t'_{2} \\ \psi_2 \end{pmatrix}
\end{align}
 which are related via:
\begin{align}
	Q_1\ \overset{T}\longleftrightarrow\ -\Sigma_0 Q_2\ ,
	\qquad\qquad \textrm{with}\quad
	\psi_i &= -i\sigma_2 \begin{pmatrix} t_i \\ b_i \end{pmatrix}\ .
\end{align}
Then, the $T$-parity invariant Lagrangian related to the top quark Yukawa interaction is given by:
\begin{align}
	\label{lagrangian}
	\mathcal{L}_{top} = -\frac{\lambda_1 f}{2\sqrt{2}} \epsilon_{ijk}\epsilon_{xy} [(\overline{Q_1})_i\Sigma_{jx}\Sigma_{ky}-(\overline{Q_2}\Sigma_0)_i\tilde{\Sigma}_{jx}\tilde{\Sigma}_{ky}] u^3_{R} - \lambda_2 f (\overline{t'}_{1} t'_{1R} + \overline{t'}_{2} t'_{2R}) + h.c.
\end{align}
where $ \lambda_1 $ and $ \lambda_2 $ are two dimensionless top-quark Yukawa couplings, $\epsilon_{ijk}$ and $\epsilon_{xy}$ are the antisymmetric tensors with $i,j,k =1, 2, 3$ and $x,y = 4, 5$, and
\begin{align}
	\tilde{\Sigma} \equiv T[\Sigma] = \Sigma_0 \Omega \Sigma^{\dagger} \Omega\Sigma_0\ , 
	\qquad\qquad \textrm{with} \quad
	\Omega = \rm diag(1,1,-1,1,1)\ .
\end{align} 

After the symmetry breaking, we can get the masses of top quark and its partners from the Lagrangian in Eq.(\ref{lagrangian}) and parameterize them at $ \mathcal{O}(v^2/f^2) $ in the following form:
\begin{align}
	m_t&=\frac{\lambda_2 v R}{\sqrt{1+R^2}} \left[ 1 + \frac{v^2}{f^2}
	\left( -\frac{1}{3} + \frac{1}{2} \frac{R^2}{(1+R^2)^2} \right)\right]\nonumber \\
	m_{T_{+}}&=\frac{f}{v}\frac{m_{t}(1+R^2)}{R}\left[1+\frac{v^{2}}{f^{2}}\left(\frac{1}{3}-\frac{R^2}{(1+R^2)^2}\right)\right] \nonumber \\
	m_{T_{-}}&=\frac{f}{v}\frac{m_{t}\sqrt{1+R^2}}{R}\left[1+\frac{v^{2}}{f^{2}}\left(\frac{1}{3}-\frac{1}{2}\frac{R^2}{(1+R^2)^2}\right)\right]\label{Tmass}
\end{align}
where $R$ is defined as $ R=\lambda_1/\lambda_2 $.

\begin{figure}[ht]
	\centering
	\begin{subfigure}{0.48\linewidth}
		\includegraphics[width=\linewidth]{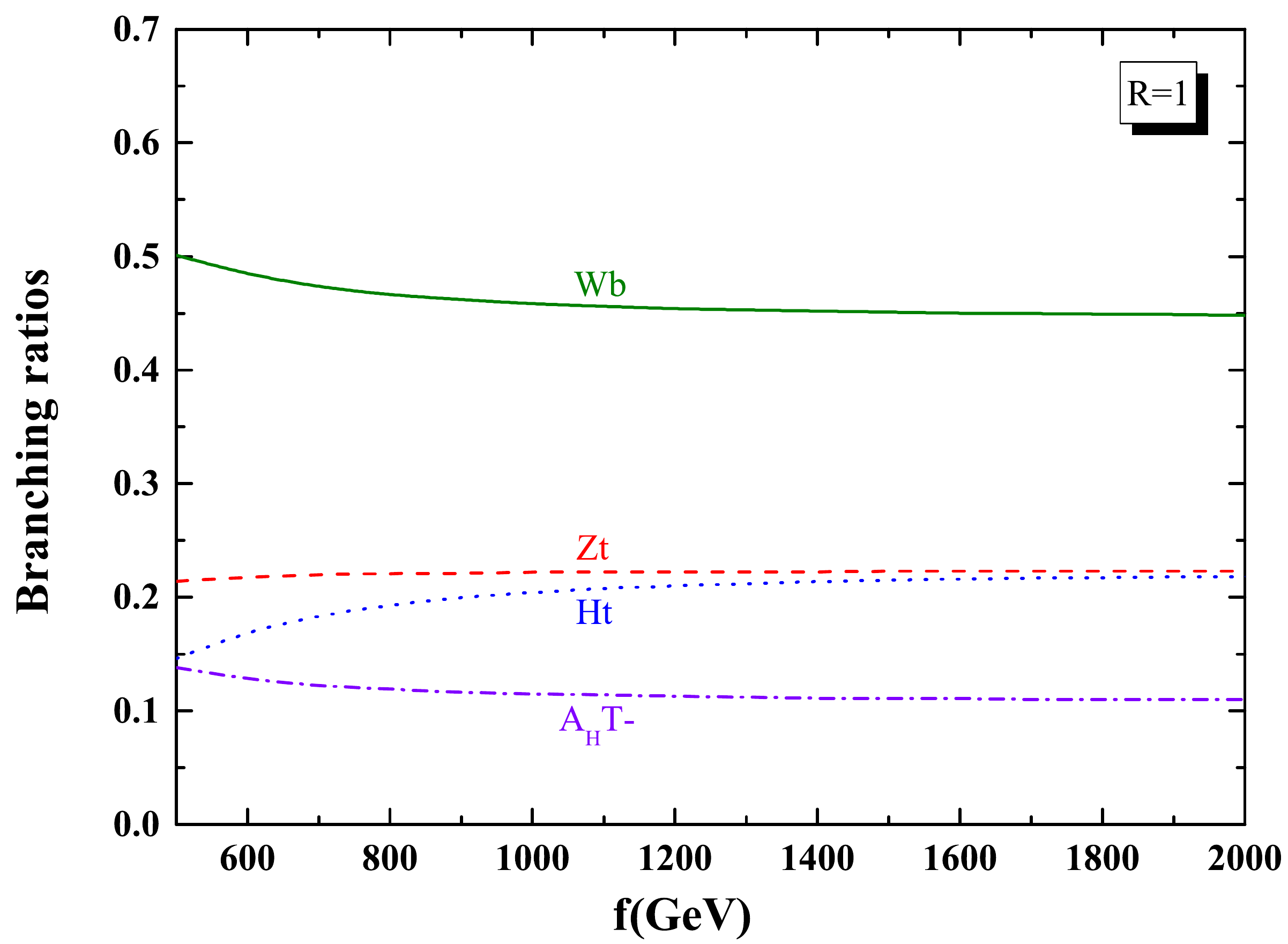}
	\end{subfigure}
	\begin{subfigure}{0.46\linewidth}
		\includegraphics[width=\linewidth]{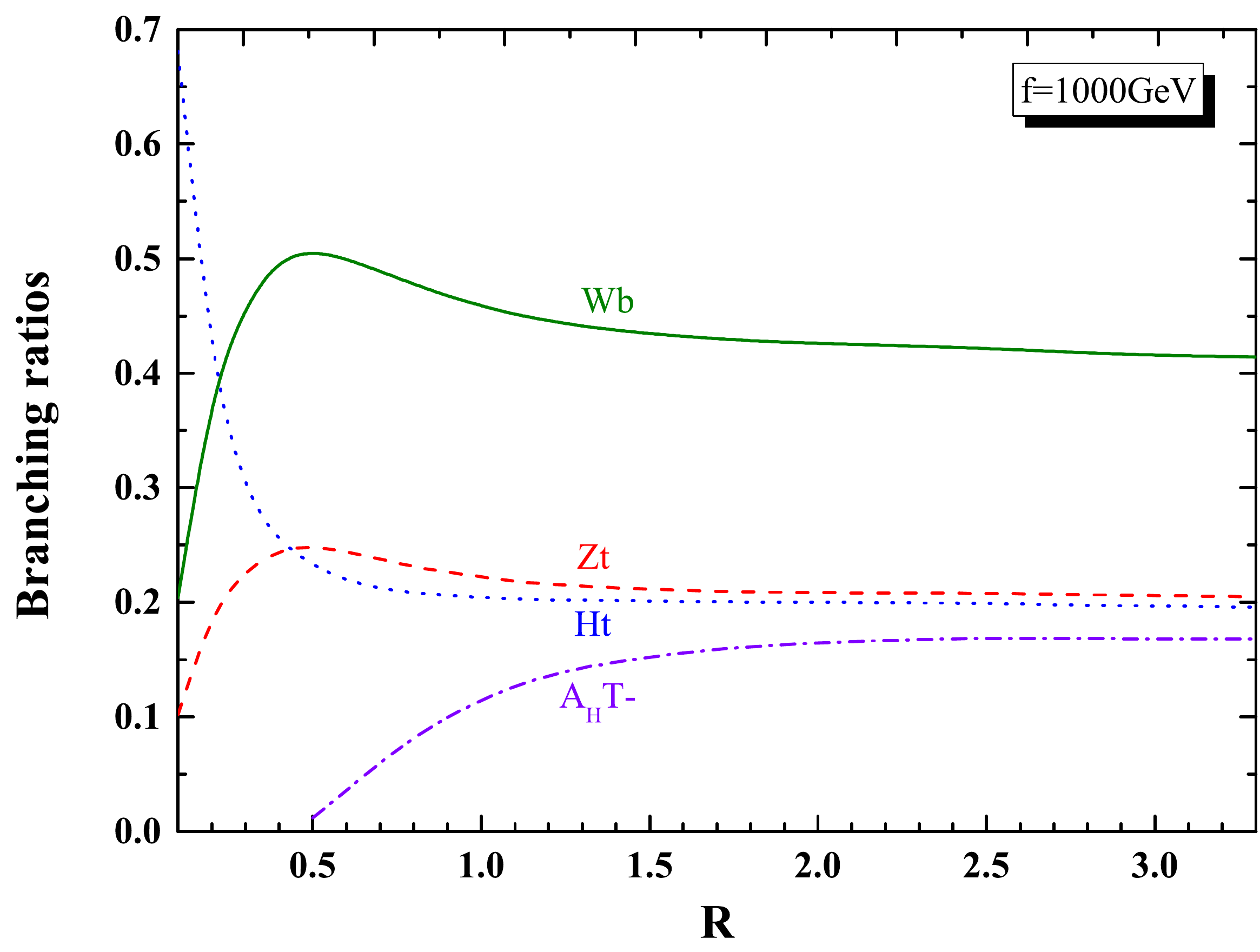}
	\end{subfigure}
	\caption{The branching ratios of $ T_+ $ as a function of the scale $f$ for $R = 1$ (left) and as a function
		of the ratio $R$ for $f = 1000$ GeV (right).}
	\label{fig:tdecay}
\end{figure}

Due to the conservation of T parity, the vector-like top partner $ T_+ $ is the only new particle that can be singly produced, which means all other new particles have to be pair produced in this model. Apart from the usual decay channels: $ T_+ \rightarrow b W^+ $, $ T_+ \rightarrow t Z $ and $ T_+ \rightarrow t h $, the $ T_+ $ has an additional decay channel $ T_+ \rightarrow T_- A_H $ so that it has richer phenomenology. \add{For clarity, we show the branching ratios of these four decay channels as a function of the scale $f$ (left) and as a function
of the ratio $R$ (right) in Fig.\ref{fig:tdecay}. We can see that the additional decay channel $ T_+ \rightarrow T_- A_H $ has a weak dependence on the scale $f$ and is mainly determined by the ratio $R$. It is kinematically opened only for $R>0.5$ when $f=1000$GeV and will help to detect the effect of the LHT model.}

Now, let's go back to the fine-tuning problem, which is the initial driving force of the LHT model. The fine-tuning problem can be quantified by the definition in Ref.\cite{fine-tuning}, the form of the fine-tuning measure is the ratio of the experimentally measured Higgs mass squared parameter($ \mu_{obs}^2 = m_h^2/2$) and the absolute value of the radiative corrections from the top partners to the Higgs quadratic operator($ \delta \mu^2 $):
\begin{align}
    \Delta = \frac{\mu_{obs}^2}{| \delta \mu^2 |}, \qquad \delta \mu^2 = -\frac{3\lambda_t m_{T_{+}}^2}{8\pi^2}\log\frac{\Lambda}{m_{T_{+}}^2}
    \label{fine-tuning}
\end{align}
Here $ \lambda_t $ is the SM top Yukawa coupling and $ \Lambda = 4\pi f $ is the cut-off scale of the LHT model. According to this definition, a smaller value means a severer fine-tuning.

\section{Single top partner production at the \texorpdfstring{$ep$}{Lg} colliders}

\begin{figure}[!ht]
	\centering
	\includegraphics[width=0.5\textwidth]{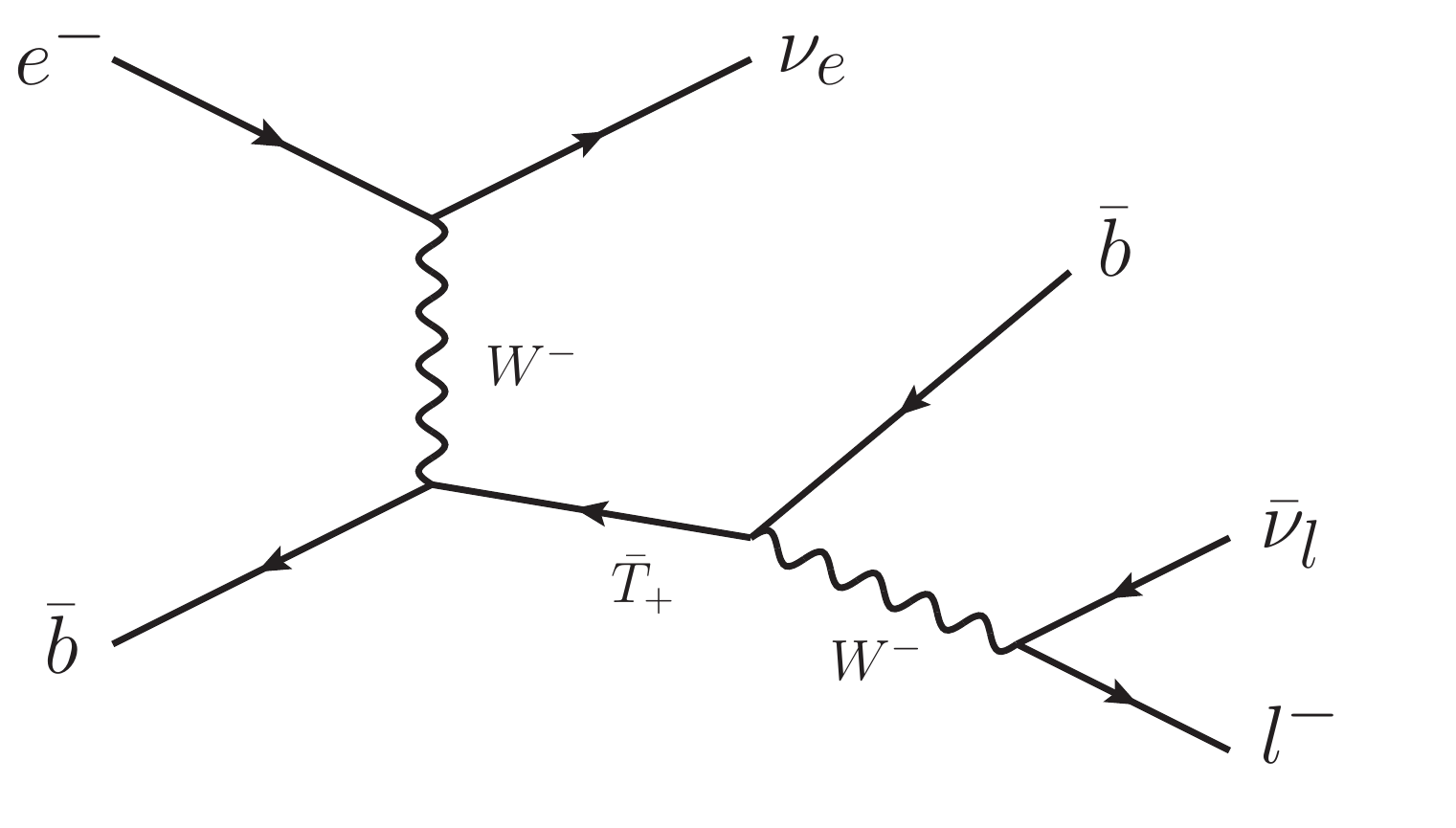}
	\caption{Single top partner production and decay to $ Wb $ channel at the $ep$ collision.}
	\label{FeynDiagram}
\end{figure}

At the $ep$ colliders, the dominant way to produce the $ T_+ $ is through $t$-channel by exchanging a $W$ boson, as shown in Fig.\ref{FeynDiagram}. The new coupling vertex $T_+ Wb$ related to this process is given by:
\begin{align}
\bar{T}_+ W^{+\mu} b: \quad & \frac{ig}{\sqrt{2}} (V_{\rm CKM})_{tb} \frac{R^2}{1+R^2} \frac{v}{f} [1+\frac{v^2}{f^2} d_2] \gamma^\mu P_L\\
\textrm{with:}\quad & d_2 = -\frac{5}{6}+\frac{1}{2}(\frac{R}{1+R^2})^2(R^2+4)
\end{align}
\add{where $ P_L $ is the chiral projection operator and $ (V_{\rm CKM})_{tb} $ is one of the CKM elements. The $ (V_{\rm CKM})_{tb} $ will not to be unit if one assumes that there is minimal flavor violation in the LHT model}\cite{DAmbrosio:2002vsn}. Here, we set $ (V_{\rm CKM})_{tb}=1$.

This process depends only on two free parameters closely related to the $ T_+ $, that is the scale $ f $ and the ratio $ R $. \add{In our previous work\cite{Wu:2016rwz}, we have considered the limits on the LHT model from the dark matter direct detections including LUX, PandaX-II and XENON1T. In order to produce the correct relic abundance, it requires the heavy photon $ A_H $ has to co-annihilate with the light mirror quarks or mirror leptons, which leads the Yukawa coupling $ \kappa $ of the mirror
fermions to be confined in a small region and causes the LHT model to become unnatural. Besides, the $ A_H $ will no longer serve as dark matter candidate in the case of T-parity violation by instantons. } Considering the current constraints\cite{LHT-limit}\cite{LHCLimit}, we take the loose parameter space and allow the scale $f$ to be as low as 500 GeV, and the range of the ratio $ R $ is chosen as $ R \in [0.1,3.3] $. The relevant SM input parameters are taken as follows \cite{pdg}:
\begin{align}
\nonumber m_t = 173.0{\rm ~GeV},\quad &m_{Z} =91.1876 {\rm ~GeV}, \quad m_h =125.0 {\rm ~GeV}, \\
\nonumber \sin^{2}\theta_W& = 0.231,\quad \alpha(m_Z) = 1/128.
\end{align}

\begin{figure}[!htb]
	\setlength{\abovecaptionskip}{0.cm}
	\setlength{\belowcaptionskip}{-0.cm}
	\centering
	\begin{subfigure}[t]{0.48\textwidth}
		\centering
		\includegraphics[width=\textwidth]{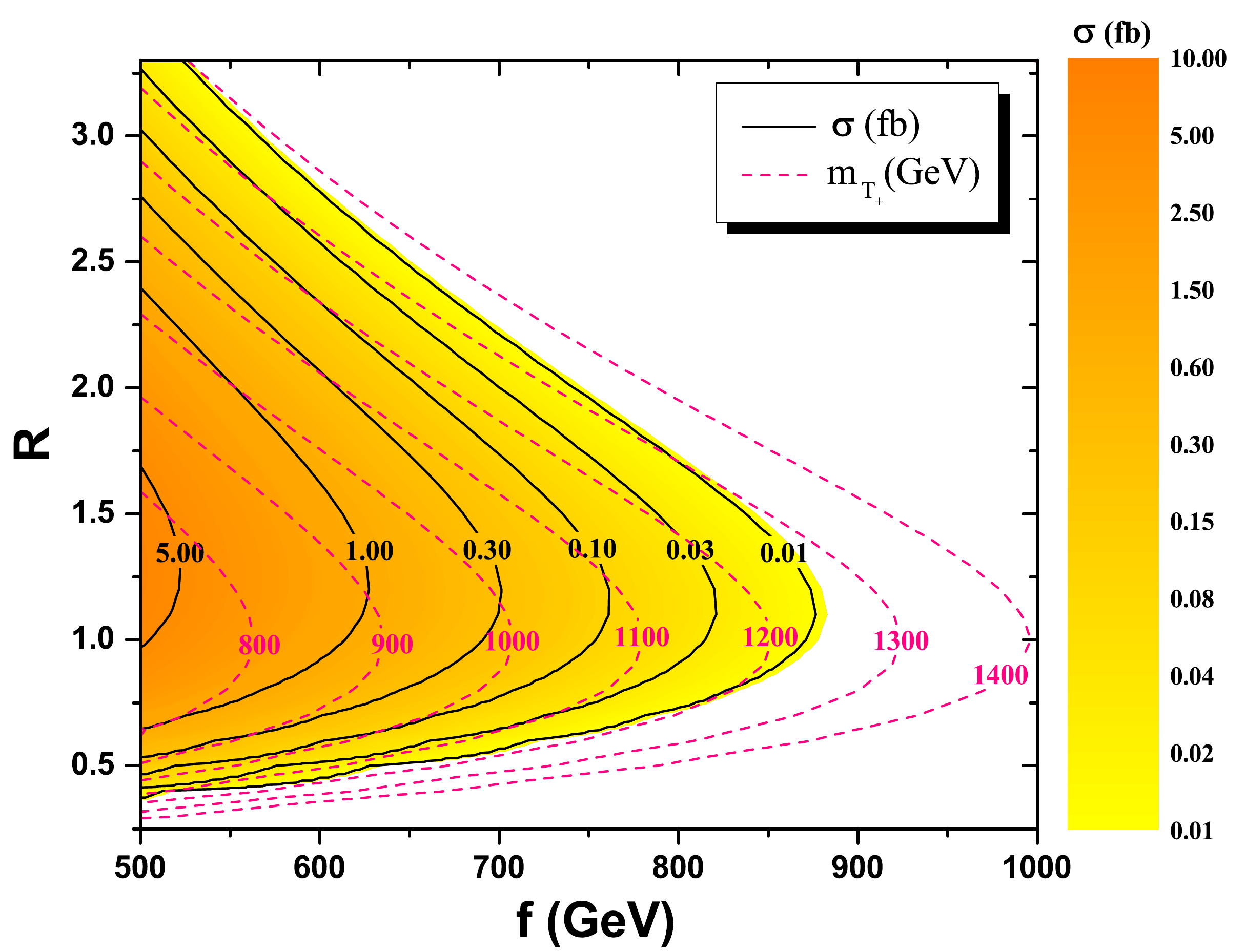}
		\caption{LHeC (unPola)}\label{production-LHeC}
	\end{subfigure}
	\begin{subfigure}[t]{0.48\textwidth}
		\centering
		\includegraphics[width=\textwidth]{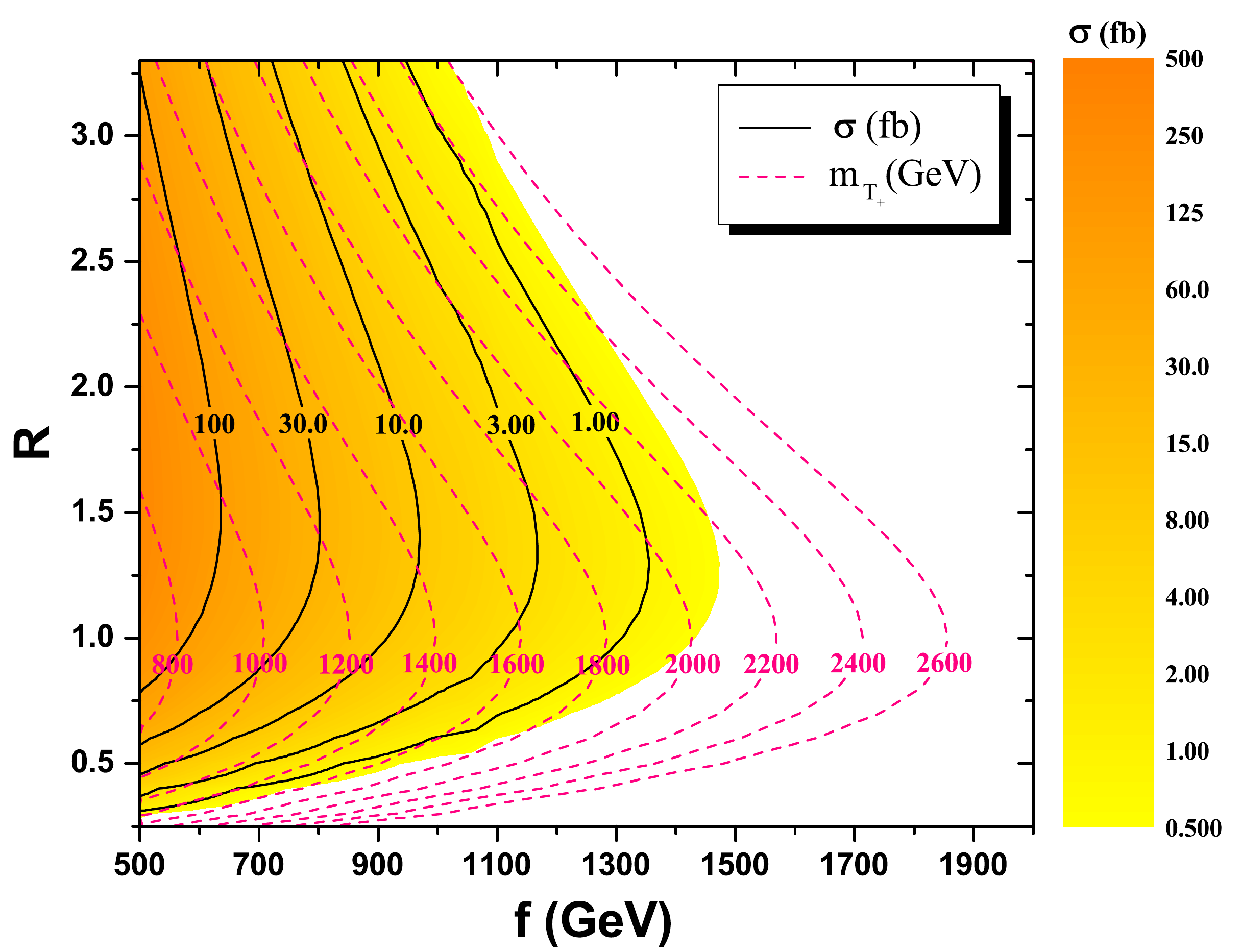}
		\caption{FCC-eh (unPola)}\label{production-FCC}
	\end{subfigure}	
	\caption{The unpolarized cross section $\sigma( e^- p \rightarrow \nu_e \bar{T}_+ ) $ distribution in the $ R-f $ plane at LHeC(a) and FCC-eh(b).}
	\label{production}
\end{figure}

For the collision energy, we choose 140 GeV $\times$ 7 TeV and 140 GeV $\times$ 50 TeV as the $e^-$ and $p$ beam energies, which correspond to LHeC($ \sqrt{s} $ = 1.98TeV) and FCC-eh($ \sqrt{s} $ = 5.29TeV), respectively. Then, we scan the parameter space with the package EasyscanHEP \cite{Easyscan} and show the unpolarized cross sections of single $ T_+ $ production at the LHeC (Fig.\ref{production}(a)) and FCC-eh(Fig.\ref{production}(b)) in the $ R-f $ plane.
We can see that the cross sections almost coincide with the curves of $ T_+ $ mass and decrease rapidly with the increase of this mass. Due to the larger center-of-mass energy, the cross section at the FCC-eh can be enhanced greatly compared to the LHeC.

As we know, the SM predicts that only the left-handed electrons participate in the charged current. So it will be helpful to consider the polarized $ e^- $ beams at the $ep$ colliders. In analogy with Eq.(1.15) in Ref.\cite{MoortgatPick:2005cw},  the cross section of this process with longitudinally-polarized $ e^- $ beams and unpolarized proton beams can be expressed as:
\begin{eqnarray}
	\sigma_{pola} \propto (1-P_{e^-})\sigma_{L}
\end{eqnarray}
where $ \sigma_{L} $ is the cross section for completely left-handed polarized $ e^- $ beams ($ P_{e^-} = -$100\%). Thus, the ratio of the polarized cross section to the unpolarized one can be written as:
\begin{eqnarray}\label{Rsigma}
	R_{\sigma} \equiv \frac{\sigma_{pola}}{\sigma_{unpola}} = 1-P_{e^-}
\end{eqnarray}

We can see that the ratios $ R_{\sigma} $ are independent of the top partner mass $ m_{T_+}$. In the following calculation, we take $ P_{e^-}=-80\% $ as a reasonable polarization degree so that the polarized beams can enhance the cross sections effectively.

Considering the larger cross section, we will concentrate on the $ Wb $ channel in this work. The complete production and decay chain is shown in Fig.\ref{FeynDiagram}:
$$ e^-\ p \rightarrow \nu_e\ \bar{T}_+ (\rightarrow \bar{b}\ W^-)\rightarrow \nu_e (\bar{b}\ \nu_l\ l^- ) \rightarrow l^- + \bar{b} + \slashed E_T$$

We can see that the signal of final states is mainly composed of one charged lepton, one $b$ jet and missing energy. The dominant SM backgrounds come from the following four processes:
\begin{itemize}
	\item Background \textbf{$ t\nu $}: $ e^-\ p \rightarrow  \bar{t}\ (\rightarrow \bar{b}\ W^-) \nu_e\rightarrow l^- +  \bar{b} + \slashed E_T $
	\item Background \textbf{$Wb\nu$} : $ e^-\ p \rightarrow \  W^-(\rightarrow l^- \ \bar{\nu}_l) \ \bar{b} \nu_e \rightarrow l^- + \bar{b} + \slashed E_T $
	\item Background \textbf{$eZb$}: $ e^-\ p \rightarrow e^-\ Z(\rightarrow \nu_l \ \bar{\nu}_l) \ b (\bar{b}) \ \rightarrow e^- + b (\bar{b}) + \slashed E_T $
	\item Background \textbf{$tZ\nu$}: $ e^-\ p \rightarrow \ \bar{t}\ (\rightarrow \bar{b}\ l^-\ \bar{\nu}_l)\  Z(\rightarrow \nu_l \ \bar{\nu}_l)\ \nu_l \rightarrow l^- + \bar{b} + \slashed E_T $
\end{itemize}
where the SM top quark is on-shell produced in the Background $t\nu$ and this process is not included in the Background $ Wb\nu $.

In this work, the calculation of cross sections and generation of signal/background events are both performed by using \textsf{MadGraph5\_aMC@NLO(MG5)} \cite{mg5} with the parton distribution function(PDF) NNPDF23 \cite{nnpdf}, where the decay width of $ T_+ $ is generated by MG5 automatically. \add{Meanwhile, the Monte-Carlo(MC) generator level cuts are selected as follows:}
\begin{eqnarray}\label{basic}
\nonumber  \Delta R(i,j) >  0.4\ &,&\quad  i,j =  l, \ j  \\
\nonumber  p_{T}^l > 10 \ \text{GeV}&,&\quad  |\eta^l|<5  \\
\nonumber  p_{T}^j > 20 \ \text{GeV}&,&\quad  |\eta^j|<5
\end{eqnarray}
where $ \Delta R(i, j) = \sqrt{(\Delta \phi)^2 + (\Delta \eta)^2} $ with $ \Delta \phi $ the difference of azimuthal angle between object $i$ and $j$, meanwhile $ \Delta \eta $ the difference of pseudo-rapidity between them.

Next, we let the \add{parton-level} events go through the \textsf{PYTHIA} \cite{PYTHIA} for the parton shower and hadronization. Then the \textsf{Delphes} \cite{Delphes} is used for the fast detector simulation, where the $anti-k_t$ algorithm \cite{anti-kt} in the delphes card of FCC-eh collider is chosen for clustering the jets with distance parameter $ \Delta R = 0.4 $. 
\add{Finally}, the reconstructed-level events derived from the above process are used to do the kinematic and cut-based analysis by \textsf{MadAnalysis5} \cite{madanalysis}.

\subsection{Detection at LHeC with \texorpdfstring{$ \sqrt{s} $}{Lg} = 1.98TeV}

In this section, we will use \add{MC} simulation to analyze the detection sensitivity of the single $ T_+ $ production at the LHeC through the channel as shown in Fig.\ref{FeynDiagram}.

\begin{figure}[!htb]
    \setlength{\abovecaptionskip}{0.cm}
    \setlength{\belowcaptionskip}{-0.cm}
    \includegraphics[width=0.45\textwidth]{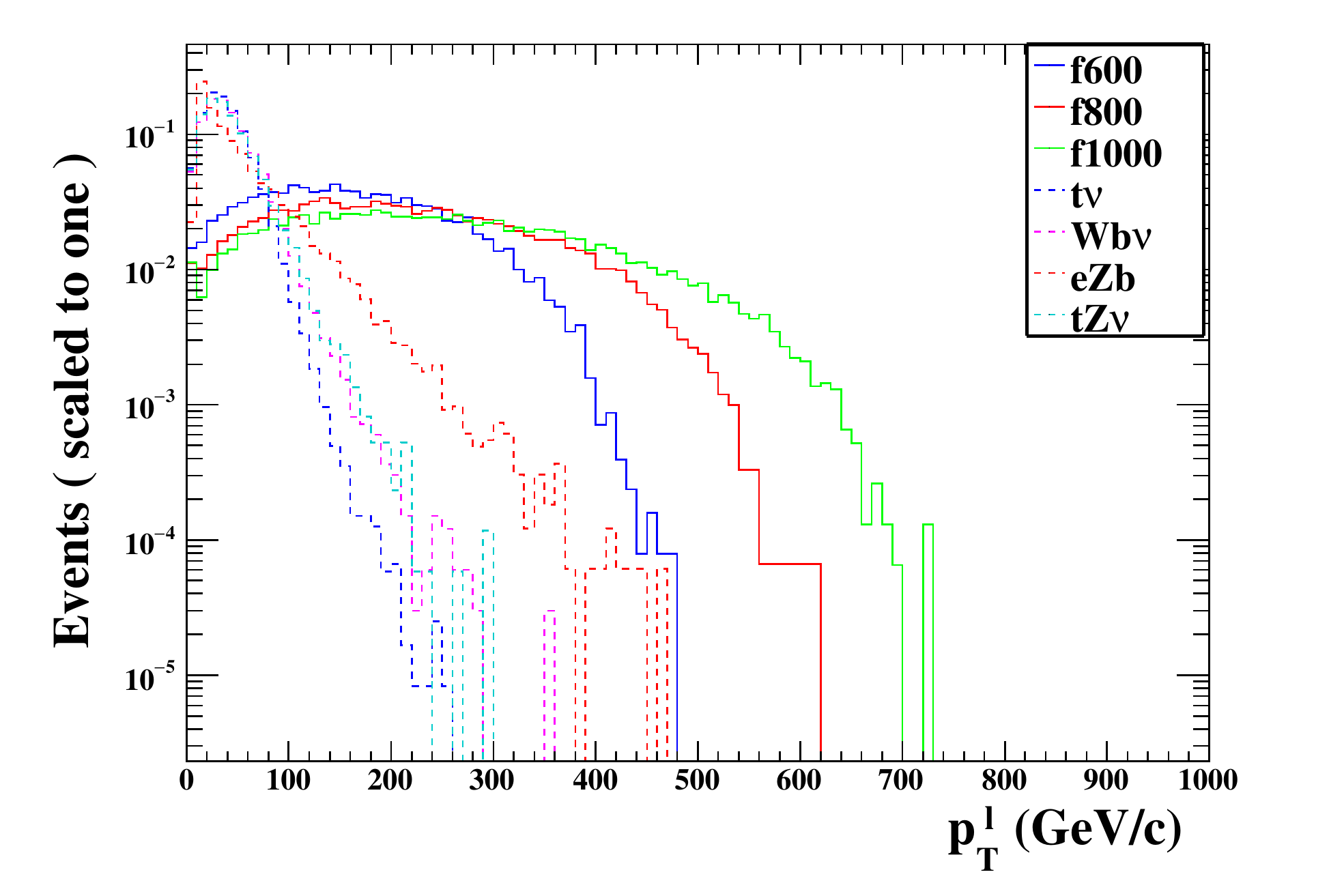}
    \includegraphics[width=0.45\textwidth]{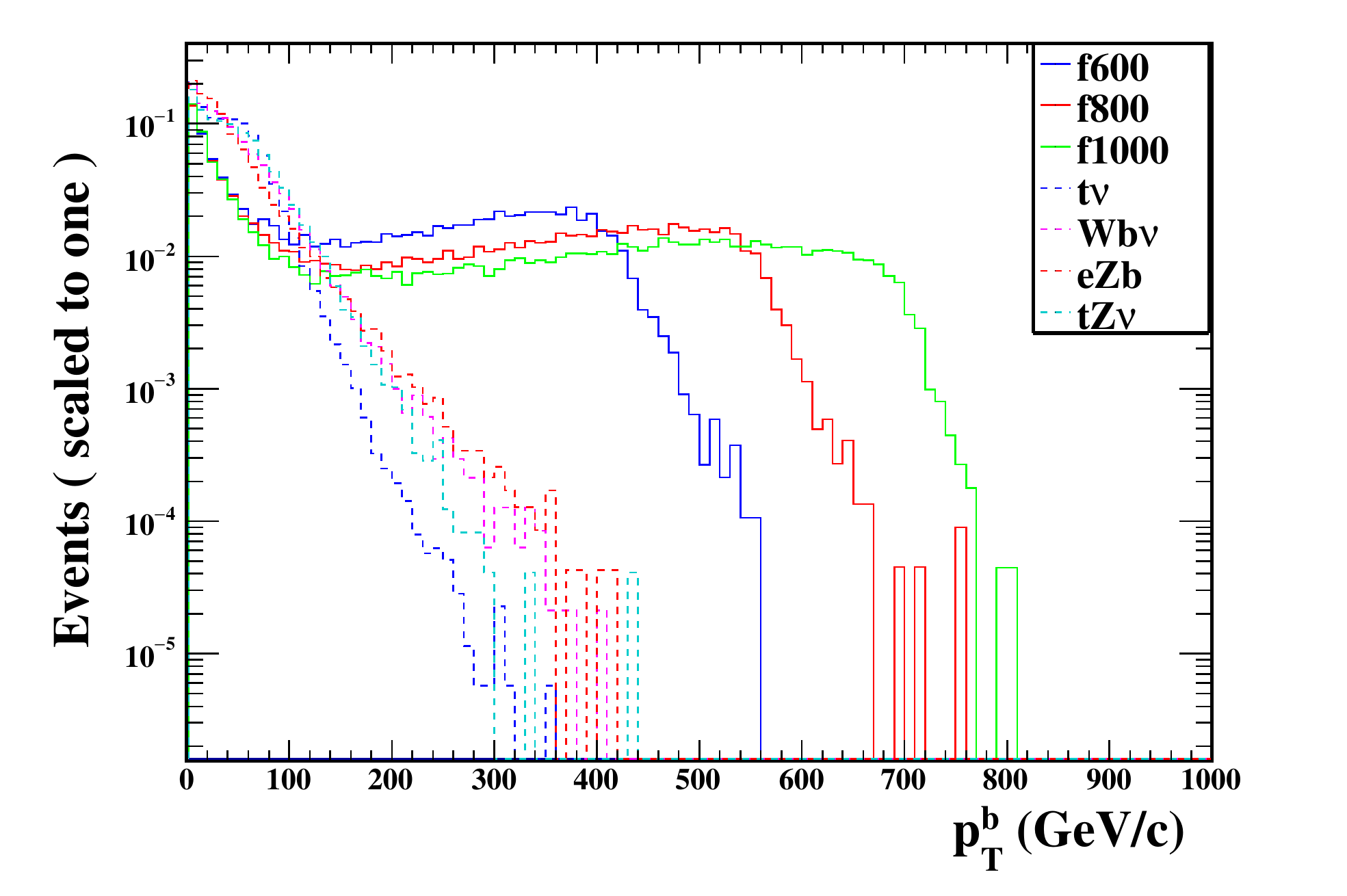}
    \includegraphics[width=0.45\textwidth]{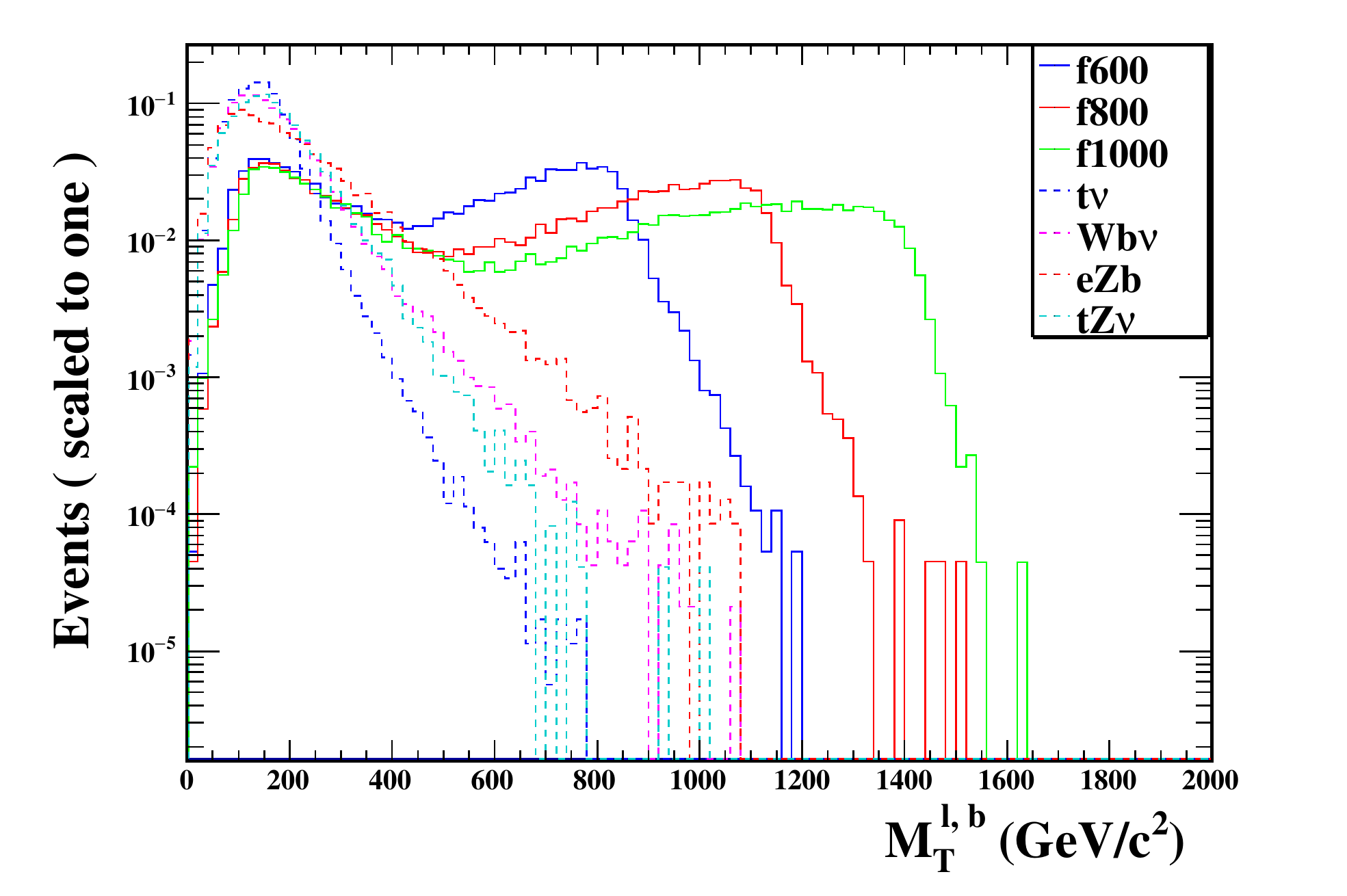}
    \includegraphics[width=0.45\textwidth]{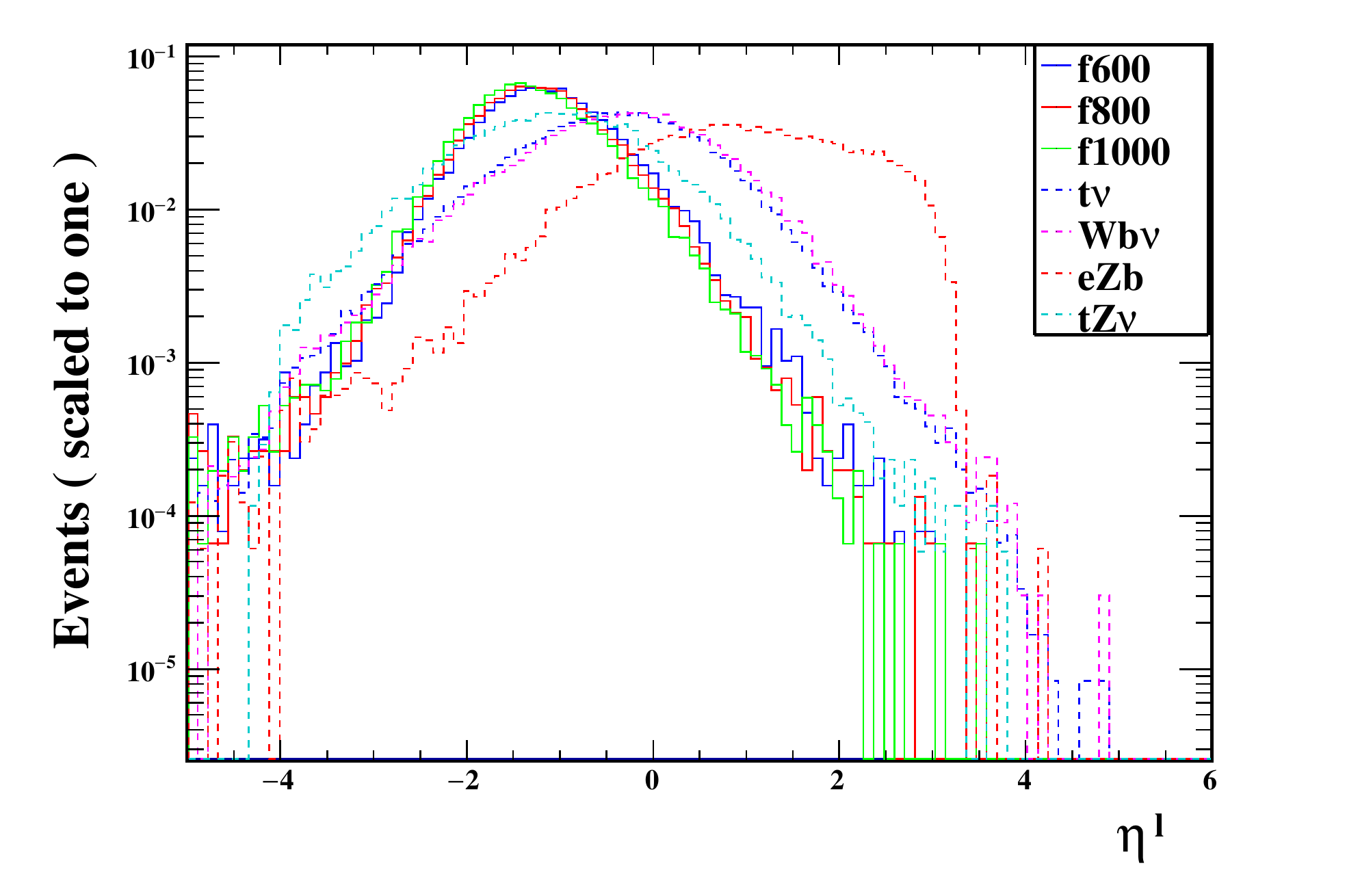}
    \includegraphics[width=0.45\textwidth]{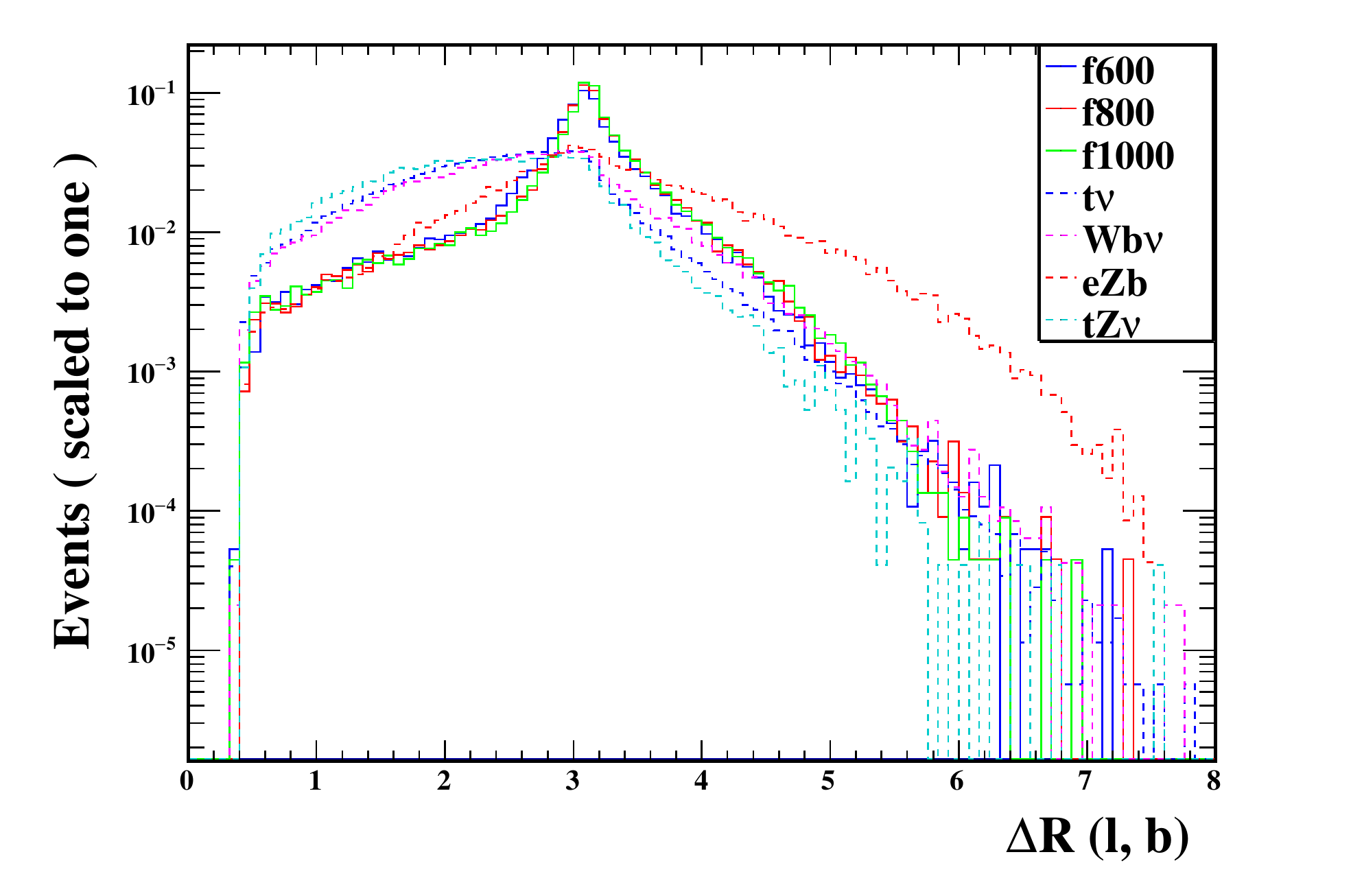}
    \caption{\label{distribution-1} The normalized distributions of $ p_T^l,\ p_T^b,\ \eta^l,\ M_T^{l,b} $ and $ \Delta R(l,b)  $ for signals and backgrounds at LHeC with unpolarized $ e^- $ beams. }
\end{figure}

Considering the blurring effect of the detector, we require that the signal contains one charged lepton(only for $ e,\mu $) and at least one $b$ jet. In order to suppress the background more effectively, some other cuts need to be taken. Since the signal leptons and $b$ quarks are derived from the decay of a heavy top partners, they should have a large transverse momentum and $ \Delta R $. In addition, the signal transverse mass, obtained from a system comprised of the lepton, $b$ quark and the invisible transverse momentum of the event \cite{MA5Recasting}, should have a peak around the $ T_+ $ mass. Besides, due to the asymmetry of initial beam energies, the pseudo-rapidity will be a very good observable to distinguish the leptons whether they come from the scattering of initial states or not. \add{Here we take the electron beam along the positive direction of the $z$ axis and the proton beam along the opposite direction.} \add{ This will lead the pseudo-rapidity of most final electrons from the scattering of initial states to positive values, and most final leptons from the decay of heavy particles to negative values due to the motion of the center of mass system caused by the large proton momentum.} The normalized distributions of these observables are shown in Fig.\ref{distribution-1} for three benchmark points $ f = 600,\ 800,\ 1000 $GeV and $ R=1 $ (corresponding to $ m_{T_+} \approx 840,\ 1120, 1400 $GeV). We find that these distributions for polarized $ e^- $ beams are roughly the same as the unpolarized ones, so only take the unpolarized case for example. According to the above analysis, we choose the specific \add{analysis} cuts as follows:
\begin{itemize}
	\item Cut1: $ N(l)=1 , N(b)\ge 1 $
	\item Cut2: $ p_T^l \ge 60 {\rm ~GeV} ,~ p_T^b \ge 200 {\rm ~GeV} $
	\item Cut3: $ M_T^{l,b} \ge 450 {\rm ~GeV} $
	\item Cut4: $ \eta^l < -0.5 $
	\item Cut5: $ \Delta R(l,b) \ge 2.0 $
\end{itemize}

We summarize the cut flows of the signals and backgrounds for unpolarized (in parenthesis) and polarized cases at LHeC in Table.\ref{efficiency-1}, in which the cross sections (in units of $10^{-3}~\rm{fb} $) can be understood as the event number with $ L=1000~\rm{fb}^{-1} $ of the signals and the backgrounds. We can find that the polarized cross section is about 1.8 times larger than the unpolarized case. After imposing the above selection cuts, we can see that the relevant backgrounds are suppressed effectively while the signal still have a relatively good efficiency. For the three benchmarks, the cut efficiencies can reach (30.2\%) 30.4\%, (35.8\%) 35.3\% and (37.1\%) 37.2\% for the (un)polarized case, respectively. In the next calculation, we will choose the conservative cut efficiency (30.2\%) 30.4\% for all the (un)polarized signal parameter points.

\begin{table}[h]
	\begin{tiny}
	\caption{Cut flows of the signals and backgrounds at LHeC with unpolarized (in parenthesis) and polarized $ e^- $ beams for the three signal benchmark points $m_{T_{+}}\approx 840, 1120, 1400 $ GeV.}
	\begin{tabular}{c|c|c|c|c|c|c|c}
		\hline \hline
		& \multicolumn{3}{c|}{Signal ($ \times 10^{-3} $fb)}& \multicolumn{4}{c} { Background ($ \times 10^{-3} $fb) } \\
		\hline
		 & 840GeV & 1120GeV & 1400GeV & $t\nu$ & $Wb\nu$ & $ eZb $ & $ tZ\nu $ \\
		\hline
		Basic Cuts 	& (129) 232 & (5.04) 9.05 & (0.095) 0.170 & (1.20E6) 2.17E6 & (18378) 33081 & (1557) 2272 & (516) 928\\
		Cut1 	& (85) 154 	& (3.32) 5.95 & (0.061) 0.110 & (7.46E5) 1.34E6 	& (11140) 19979 & (930) 1375  & (328) 592\\
		Cut2 	& (49) 89	& (2.23) 3.95 & (0.043) 0.077 & (124) 305 	& (34) 71  	& (9) 14 & (0.44) 0.69 \\
		Cut3 	& (49) 89	& (2.23) 3.95 & (0.043) 0.077 & (118) 282 	& (31) 62 	& (8) 13 & (0.40) 0.69 \\
		Cut4 	& (42) 77 	& (1.96) 3.45 & (0.038) 0.068 & (6) 6 	& (19) 42 & (7) 12 & (0) 0 \\
		Cut5 	& (39) 71 	& (1.81) 3.19 & (0.035) 0.063 & (6) 6 	& (12) 23 & (4) 5  & (0) 0 \\
		\hline
		Total Eff. & (30.2\%)30.4\% & (35.8\%)35.3\% & (37.1\%)37.2\% & (5.17E-6) 2.68E-6 & (6.38E-4) 6.94E-4  & (0.23\%)0.24\% & (0) 0 \\
		\hline \hline
	\end{tabular}
	\label{efficiency-1}
	\end{tiny}
\end{table}

The statistical significance($SS$) can be evaluated after the final cut by using the Poisson formula\cite{ss}:
\begin{equation}\label{eq:ss}
SS = \sqrt{2 L [(\sigma_S + \sigma_B)\ln(1+\frac{\sigma_S}{\sigma_B})-\sigma_S]}
\end{equation}
where $ L $ is the integrated luminosity and $ \sigma_{S}, \sigma_{B}$ are the effective cross sections after selection cuts for signal and background, respectively. 

\begin{figure}[!htb]
	\setlength{\abovecaptionskip}{0.cm}
	\centering
	\begin{subfigure}[t]{0.48\textwidth}
		\centering
		\includegraphics[width=\textwidth]{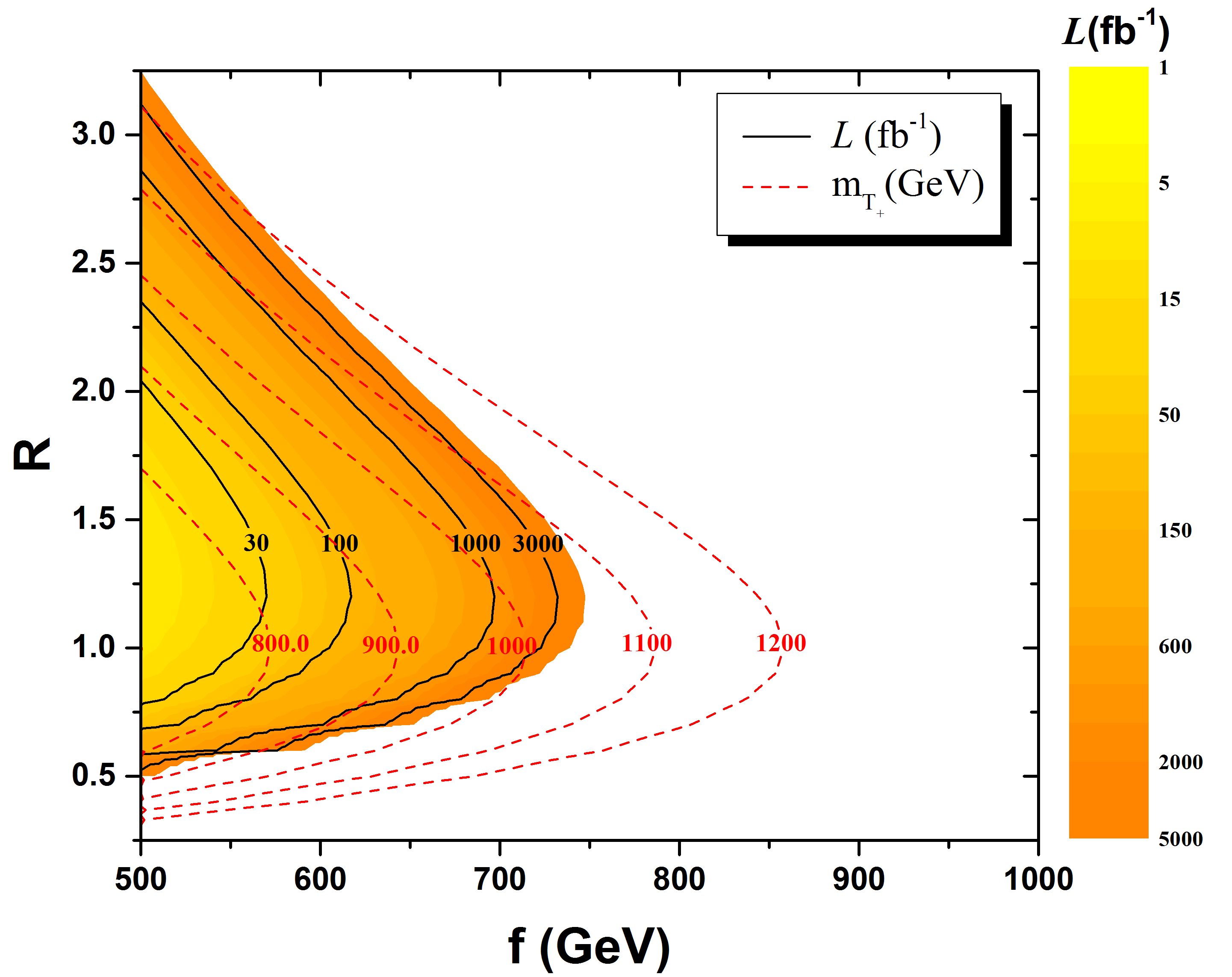}
		\caption{LHeC (unPola)}\label{LHeC-2sigma-unpola}
	\end{subfigure}
	\begin{subfigure}[t]{0.48\textwidth}
		\centering
		\includegraphics[width=\textwidth]{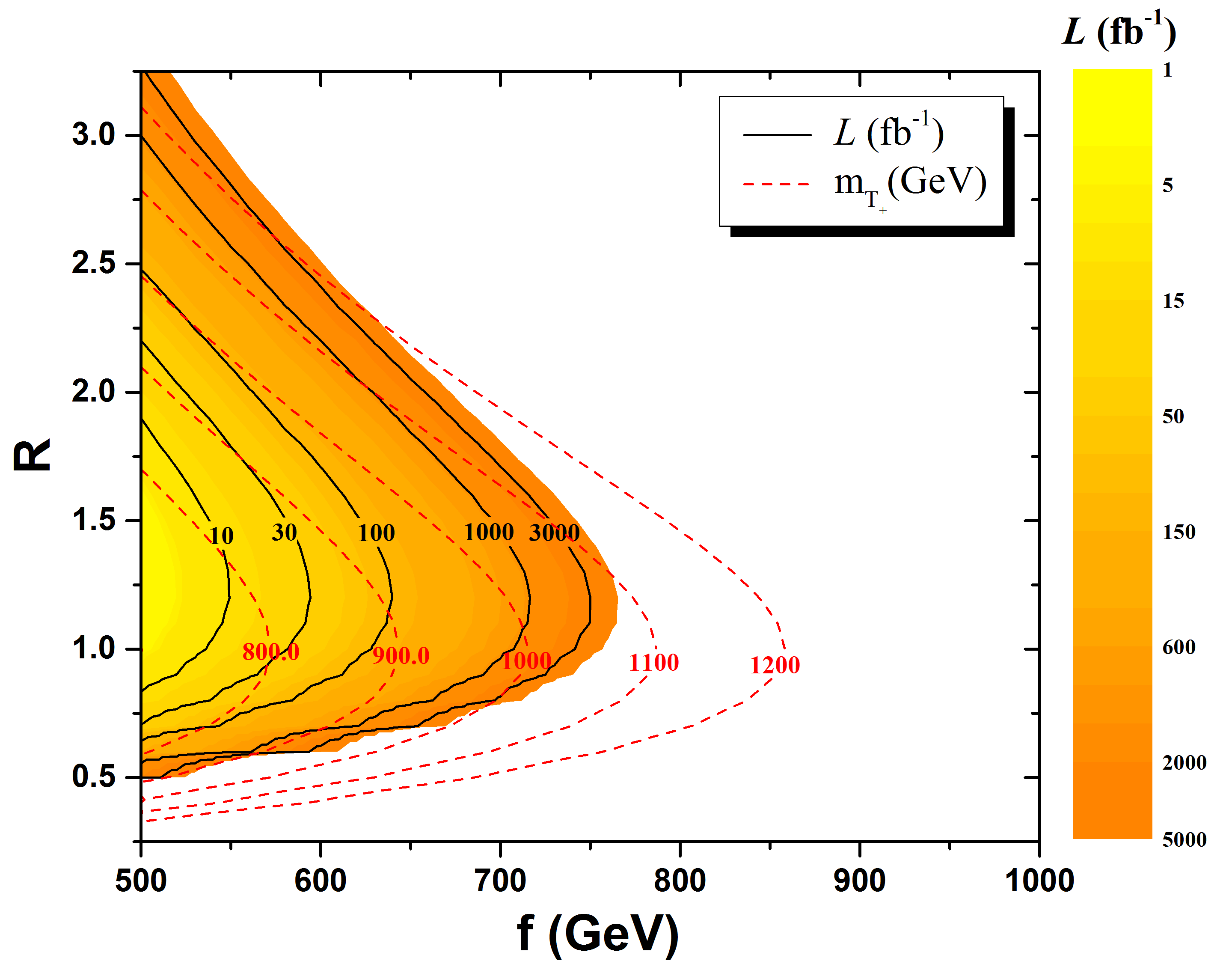}
		\caption{LHeC (Pola)}\label{LHeC-2sigma-pola}
	\end{subfigure}	
	\caption{2$ \sigma $ exclusion limit in $ R-f $ plane at the LHeC with unpolarized $ e^- $ beams(a) and polarized $ e^- $ beams(b) for $ \sqrt{s} $ = 1.98TeV. }
	\label{lhec-2sigmaLimit}
\end{figure}

\begin{figure}[!htb]
	\setlength{\abovecaptionskip}{0.cm}
	\centering
	\begin{subfigure}[t]{0.48\textwidth}
		\includegraphics[width=\textwidth]{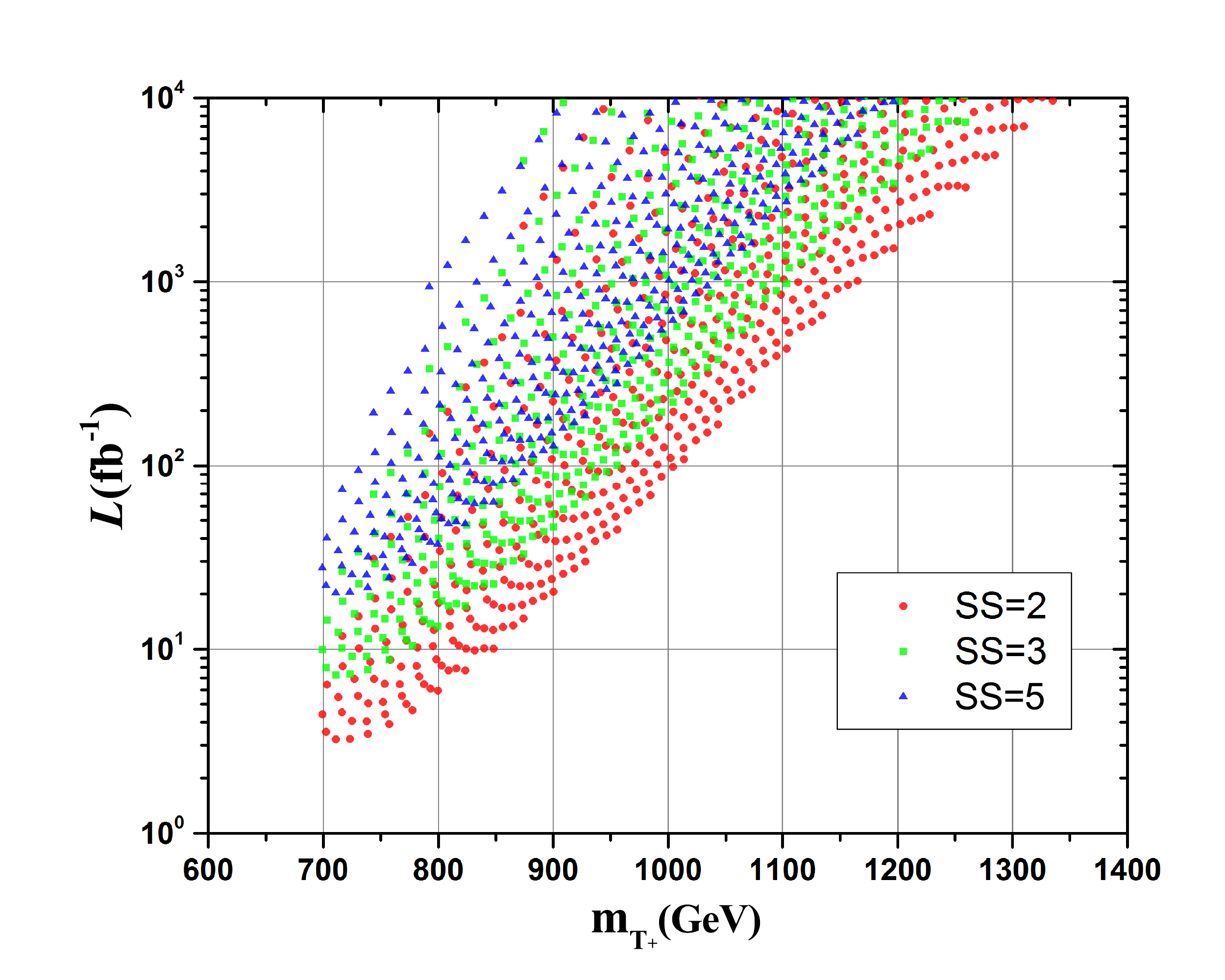}
	\end{subfigure}
	\begin{subfigure}[t]{0.48\textwidth}
		\includegraphics[width=\textwidth]{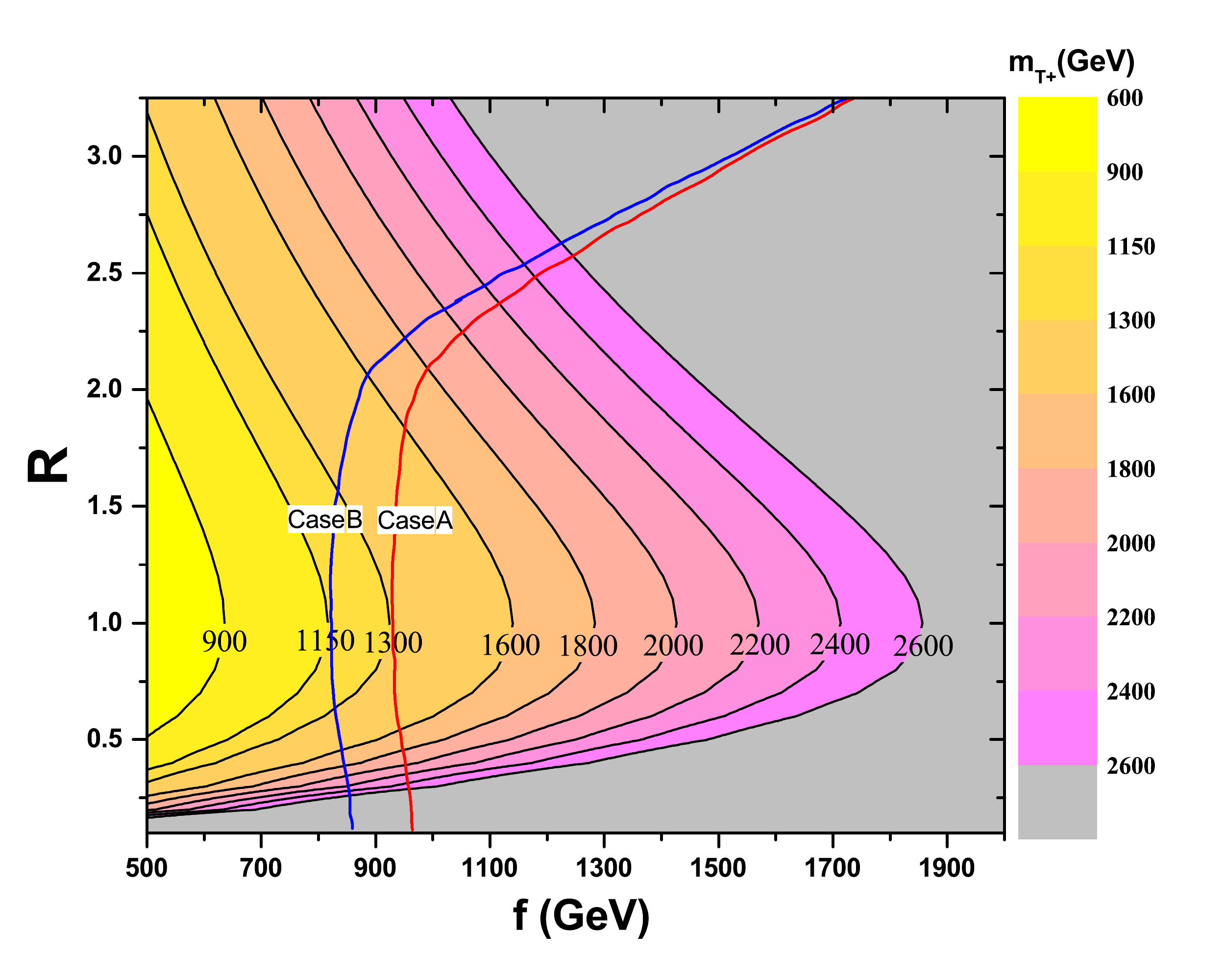}
	\end{subfigure}
	\caption{\label{ewpo-235sigmaLimit} (left) 2$\sigma$, 3$\sigma$, 5$\sigma$ samples in $ L-m_{T_{+}} $ plane at the LHeC with polarized $ e^- $ beams; (right) 2$\sigma$ limits from EWPO and Higgs data on the $m_{T_{+}}$ in the $R-f$ plane for Case A and Case B. }
\end{figure}

We show the 2$ \sigma $ exclusion limit contour in $ R-f $ plane at the LHeC with unpolarized(a) and polarized(b) $ e^- $ beams in Fig.\ref{lhec-2sigmaLimit} and find that the polarized $ e^- $ beam can test larger parameter space. \add{Furthermore, the polarization of initial electron beams can still improve the $SS$ although both the cross sections of signal and backgrounds are increased synchronously.} So, we will focus on the polarized case to study the observability of the signal. In Fig.\ref{lhec-2sigmaLimit}, we also provide the corresponding $ m_{T_{+}} $ contours indicated by the red dashed lines. According to the different integrated luminosities, the relevant contour regions associated with the $ R-f $ plane will be detected or excluded. In order to provide more information, we also show the 2$\sigma$, 3$\sigma$, 5$\sigma$ samples in $ L-m_{T_{+}} $ plane at the LHeC with polarized $ e^- $ beams in the left panel of Fig.\ref{ewpo-235sigmaLimit}.

In principle, we have to consider the limits on the LHT model from the indirect measurements. In our previous paper \cite{workcpc}, the global fit of the EWPO and the latest Higgs data have been performed. We show the $2\sigma$ exclusion limits on the $T_{+}$ mass in the right panel of Fig.\ref{ewpo-235sigmaLimit}, where the $T_{+}$ mass can be excluded up to 1300(1150) GeV for CaseA(B). The Case A and Case B denote two possible ways to construct the $T$-invariant Yukawa interactions of the down-type quarks and charged leptons \cite{caseAB}, which do not differ in the LHT collider phenomenology related to our work. Since the lower limit on $m_{T_{+}}$ in Case B is weaker than that in Case A, we will focus on Case A in the following discussions. Even if the high luminosities ($ L=1 \rm{ab}^{-1}\sim 5 \rm{ab}^{-1} $) are used, we can see that the limit from the LHeC search for the $T_{+}$ in this $Wb$ channel is still weaker than the current limit from the indirect measurements, i.e. the global fit of the EWPO and the latest Higgs data.

\subsection{Detection at FCC-eh with \texorpdfstring{$ \sqrt{s} $}{Lg} = 5.29TeV}

\begin{figure}[!htb]
    \setlength{\abovecaptionskip}{0.cm}
    \setlength{\belowcaptionskip}{-0.cm}
    \includegraphics[width=0.45\textwidth]{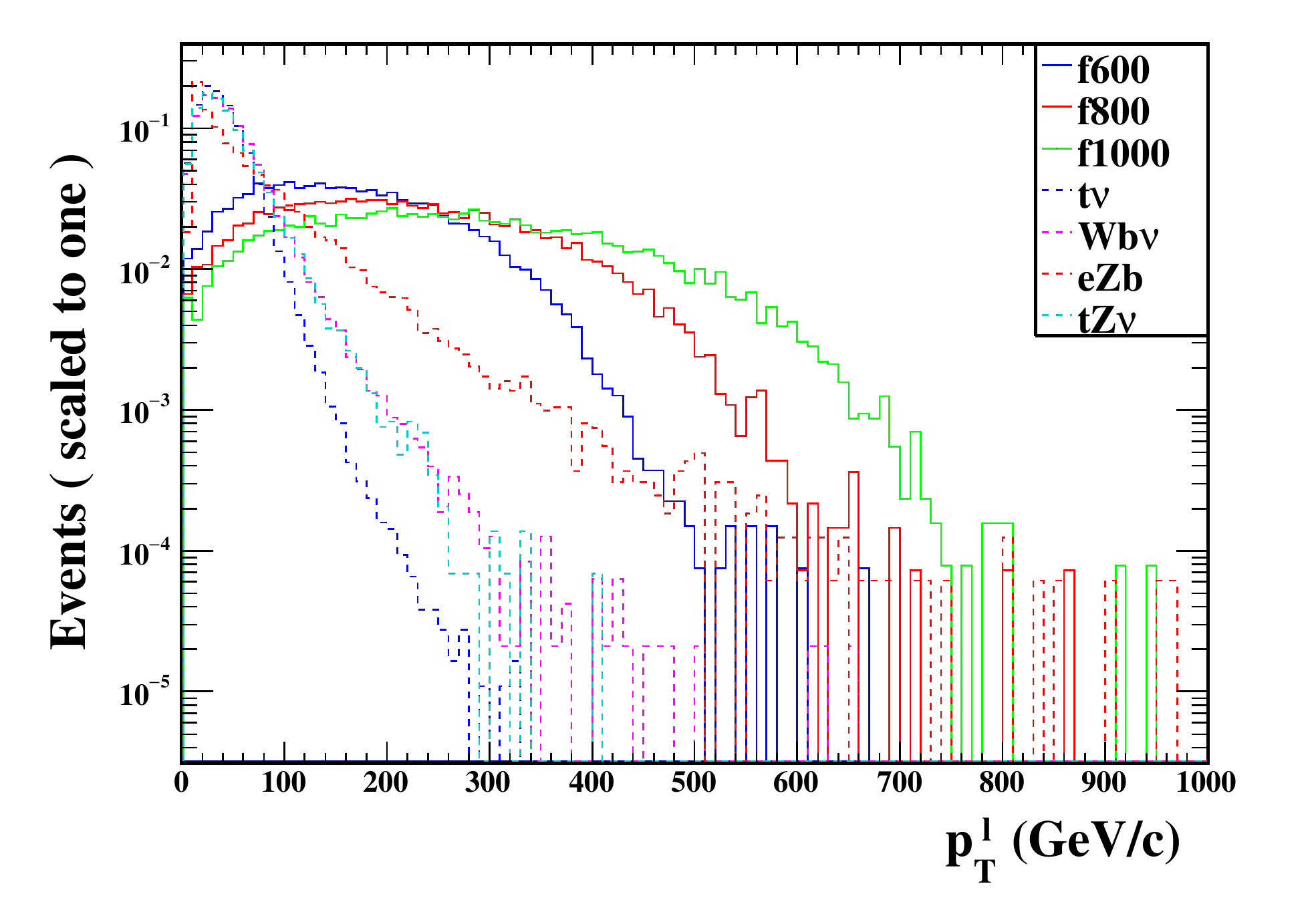}
    \includegraphics[width=0.45\textwidth]{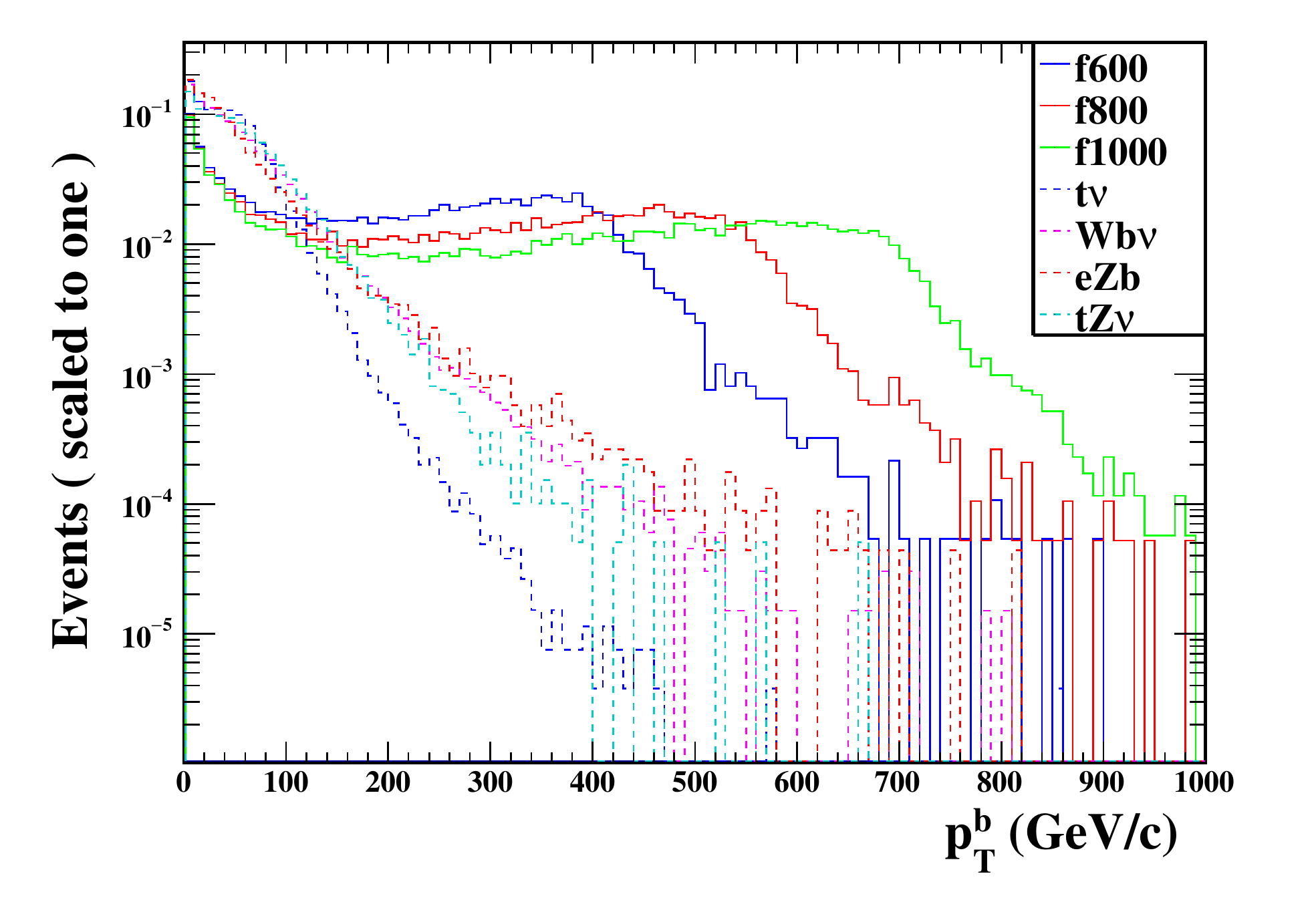}
    \includegraphics[width=0.45\textwidth]{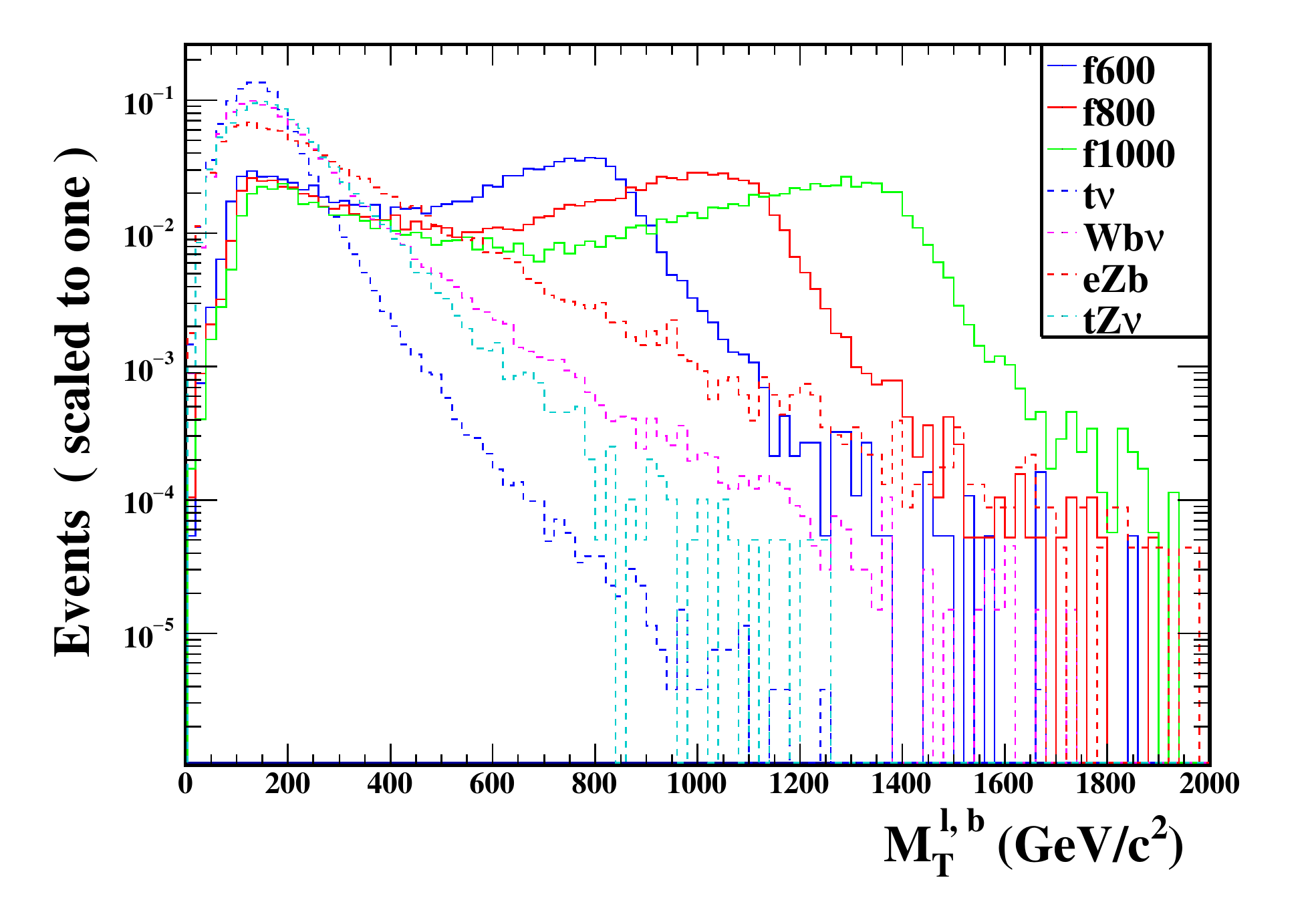}
    \includegraphics[width=0.45\textwidth]{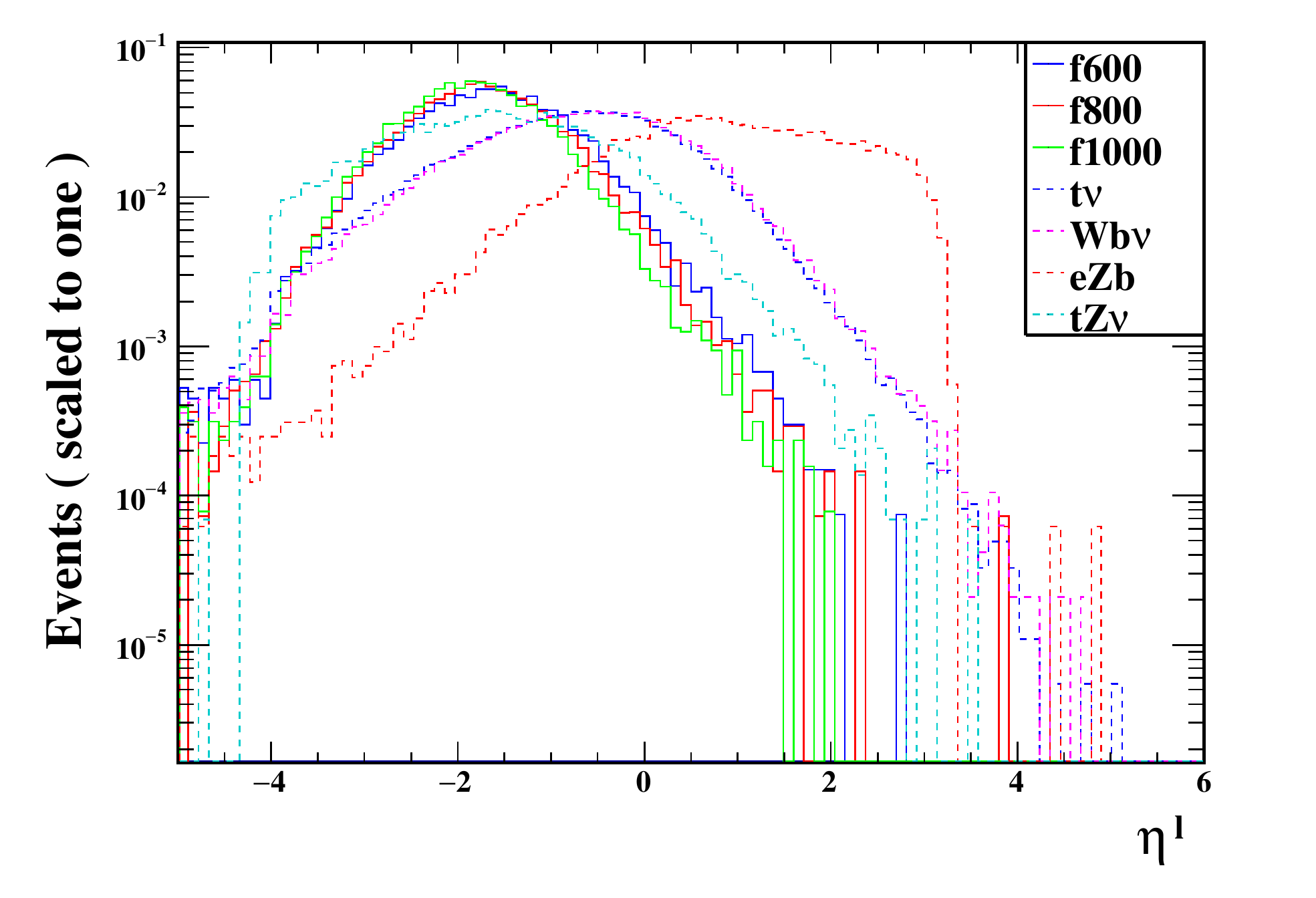}
    \includegraphics[width=0.45\textwidth]{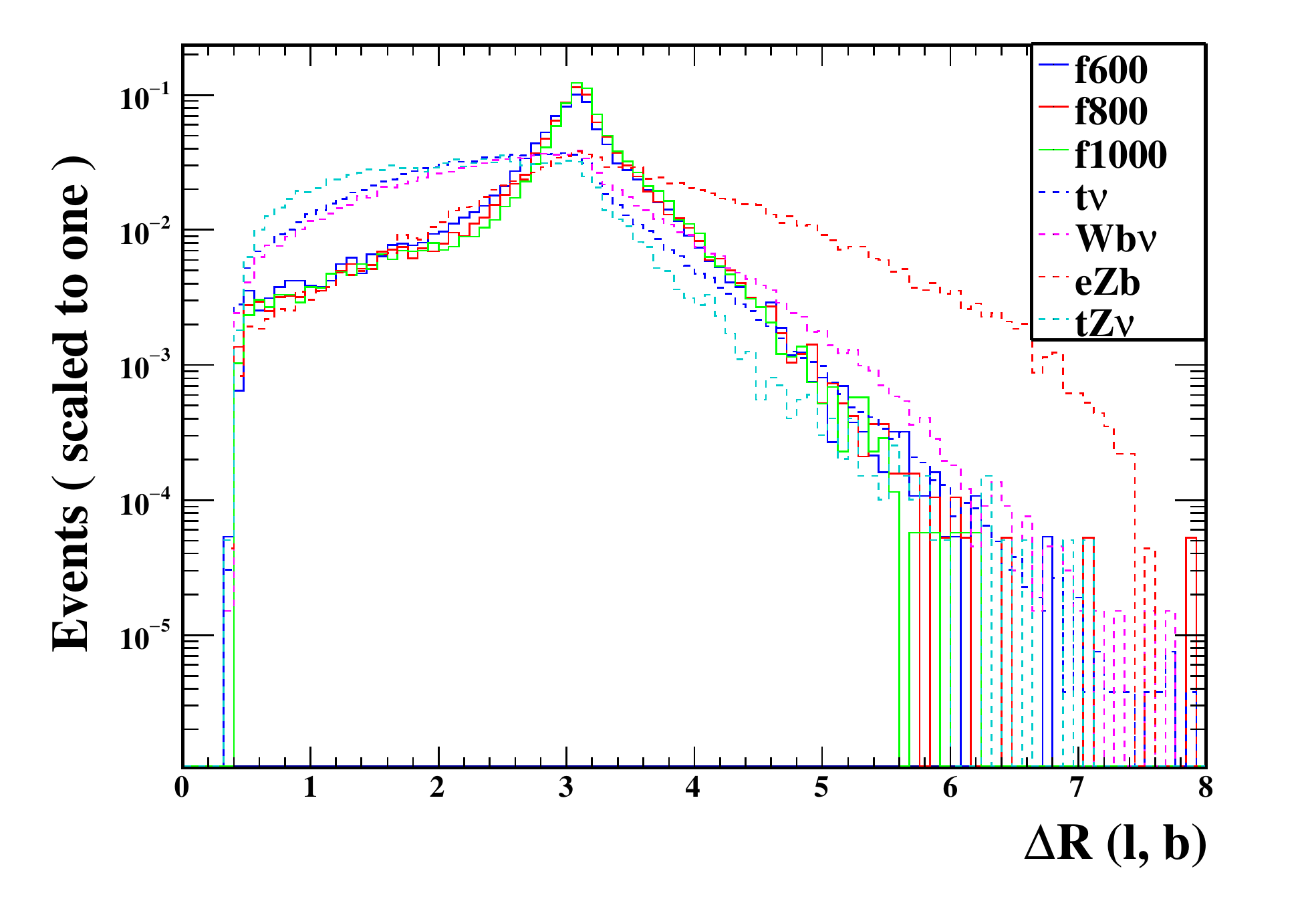}
    \caption{Same as Fig. \ref{distribution-1}, but for FCC-eh.}
    \label{distribution-2}
\end{figure}

In this section, we investigate the observability of single $ T_+ $ production at the FCC-eh with $ \sqrt{s} = 5.29$ TeV in a similar approach as the previous section. To execut the cut-based analysis, we show the same normalized kinematic distributions for the three benchmarks ($f = 600,800,1000$ GeV and $ R=1 $) in Fig.\ref{distribution-2}. We can see that \add{the distributions involving momentum show slight overall rightward shifts compared to the LHeC case, which changes the cut efficiencies almost negligibly}. So we choose the same cuts as in Sec.III(A) and the cut-flows of the signals and backgrounds for unpolarized and polarized cases are shown in Table.\ref{efficiency-3}. Here, the cut efficiencies of the (un)polarized case can reach (31.9\%) 32.4\%, (38.4\%) 38.7\% and (41.3\%) 41.0\% for three signal benchmark points, respectively. Meanwhile, the backgrounds are suppressed effectively. Likewise, we will choose the conservative cut efficiency (31.9\%) 32.4\% for all the (un)polarized signal parameter points.

\begin{table}[!htb]
	\begin{tiny}
	\caption{\small Cut flows of the signal and backgrounds at FCC-eh with unpolarized (in parentheses) and polarized $ e^- $ beam for the three signal benchmark points $m_{T_{+}}\approx 840, 1120, 1400 $ GeV.}
	\begin{tabular}{c|c|c|c|c|c|c|c}
		\hline \hline
		& \multicolumn{3}{c|}{Signal ($ \times 10^{-3} $fb)}& \multicolumn{4}{c} { Background ($ \times 10^{-3} $fb) } \\
		\hline
		 & 840GeV & 1120GeV & 1400GeV & $t\nu$ & $Wb\nu$ & $ eZb $ & $ tZ\nu $ \\
		\hline
		Basic Cuts & (8843) 1.59E4 & (2086) 3754 & (590) 1059 & (7.05E6) 1.27E7 & (1.17E5) 2.11E5 & (9123) 1.35E4 & (6695) 1.20E4\\
		Cut1 & (5489) 9956 	& (1303) 2329 & (366) 652 & (4.29E6) 7.72E6 & (6.90E4) 1.24E5 & (5464) 8216 & (3737) 6727\\
		Cut2 & (3315) 6100 	& (920) 1658 & (277) 494 & (5171) 9340 	& (882) 1662 & (189) 333 & (25) 30\\
		Cut3 & (3294) 6048 	& (915) 1652 & (276) 492 & (4627) 8612 	& (820) 1538 & (183) 326 & (24) 28\\
		Cut4 & (2918) 5346 	& (822) 1487 & (249) 444 & (118) 465  	& (545) 1038 & (164) 287 & (0.26) 1\\
		Cut5 & (2825) 5154 	&( 801) 1453 & (244) 435 & (94) 464		& (373) 754  & (78) 144  & (0.26) 1\\
		\hline
		Total Eff. 	& (31.9\%)32.4\% & (38.4\%)38.7\% & (41.3\%)41.0\% & (1.33E-5) 3.66E-5 & (0.32\%)0.36\% & (0.85\%)1.06\% & (3.85E-5) 7.69E-5\\
		\hline \hline
	\end{tabular} 
	\label{efficiency-3}
	\end{tiny}
\end{table}

\begin{figure}[!htb]
	\setlength{\abovecaptionskip}{0.cm}
	\setlength{\belowcaptionskip}{-0.cm}
	\centering
	\begin{subfigure}[t]{0.48\textwidth}
		\centering
		\includegraphics[width=\textwidth]{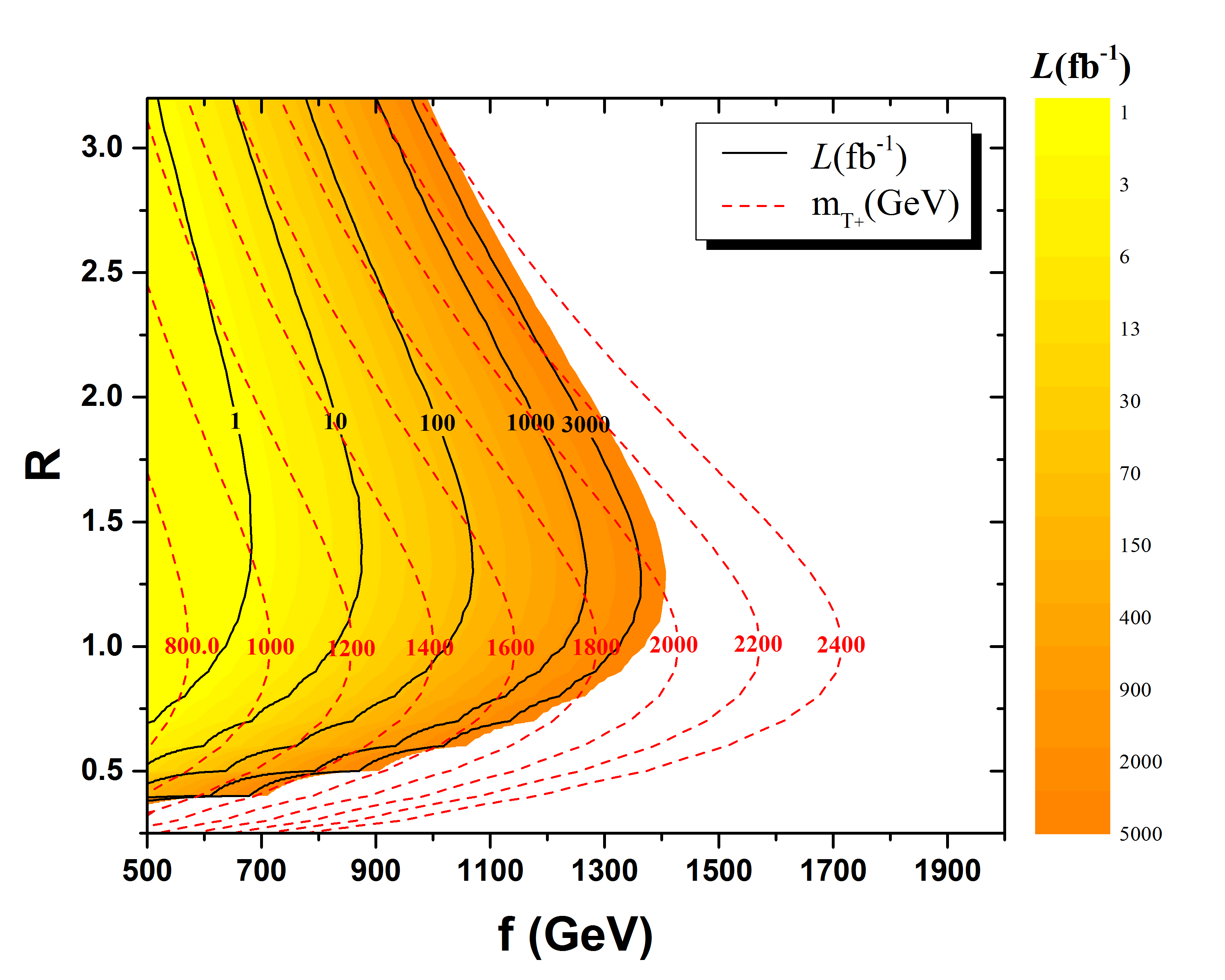}
		\caption{FCC-eh (unPola)}\label{FCC-2sigma-unpola}
	\end{subfigure}
	\begin{subfigure}[t]{0.48\textwidth}
		\centering
		\includegraphics[width=\textwidth]{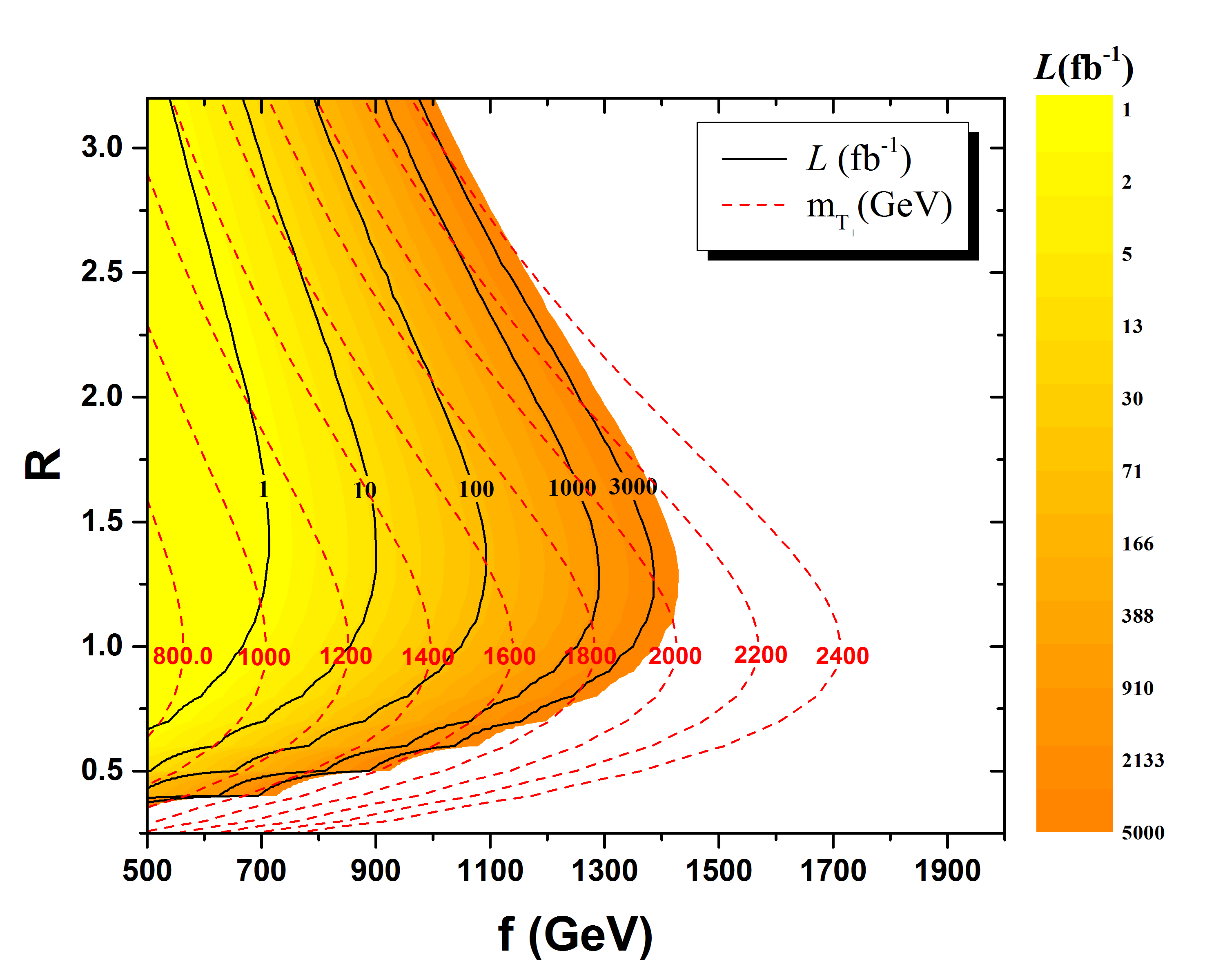}
		\caption{FCC-eh (Pola)}\label{FCC-2sigma-pola}
	\end{subfigure}	
	\caption{Same as Fig. \ref{lhec-2sigmaLimit}, but for FCC-eh.}
	\label{FCC-2sigmaLimit}
\end{figure}
In Fig.\ref{FCC-2sigmaLimit}, we show the 2$ \sigma $ exclusion limit contour in the $ R-f $ plane at the FCC-eh with unpolarized(a) and polarized(b) $ e^- $ beams. From Fig.\ref{FCC-2sigmaLimit}, we can see that the detected region at the FCC-eh is enlarged obviously with respect to the LHeC. 

Recently, the limits on the LHT model from the LHC experiments have been performed in Ref.\cite{LHCLimit}, where all the LHC available Run 2 data at 8 and
13 TeV for searches for physics beyond the SM have been exploited \add{and an extrapolation to the HL-LHC at 14TeV has also been given}. According to their conclusions, the minimum value of the scale $ f $ allowed by the LHC-13TeV experiment is $ 950 $ GeV with fixed value $ R=1 $ at 2$\sigma$ confidence level, which corresponds to the top partner mass limit $ m_{T_{+}} \ge 1336$ GeV. \add{For the HL-LHC case, the minimum value of the scale $ f $ allowed is $ 1500 $ GeV with fixed value $ R=1 $ at 2$\sigma$ confidence level, which corresponds to the top partner mass limit $ m_{T_{+}} \ge 2100$ GeV.}
Here, the value $R = 1$ corresponds to the case where minimal fine-tuning and minimal top partner mass $ m_{T_{+}} $ can be achieved so that this benchmark case can test the natural regions of the LHT parameter space. 
\begin{figure}[!htb]
	\setlength{\abovecaptionskip}{0.cm}
	\setlength{\belowcaptionskip}{-0.cm}
	\includegraphics[width=0.48\textwidth]{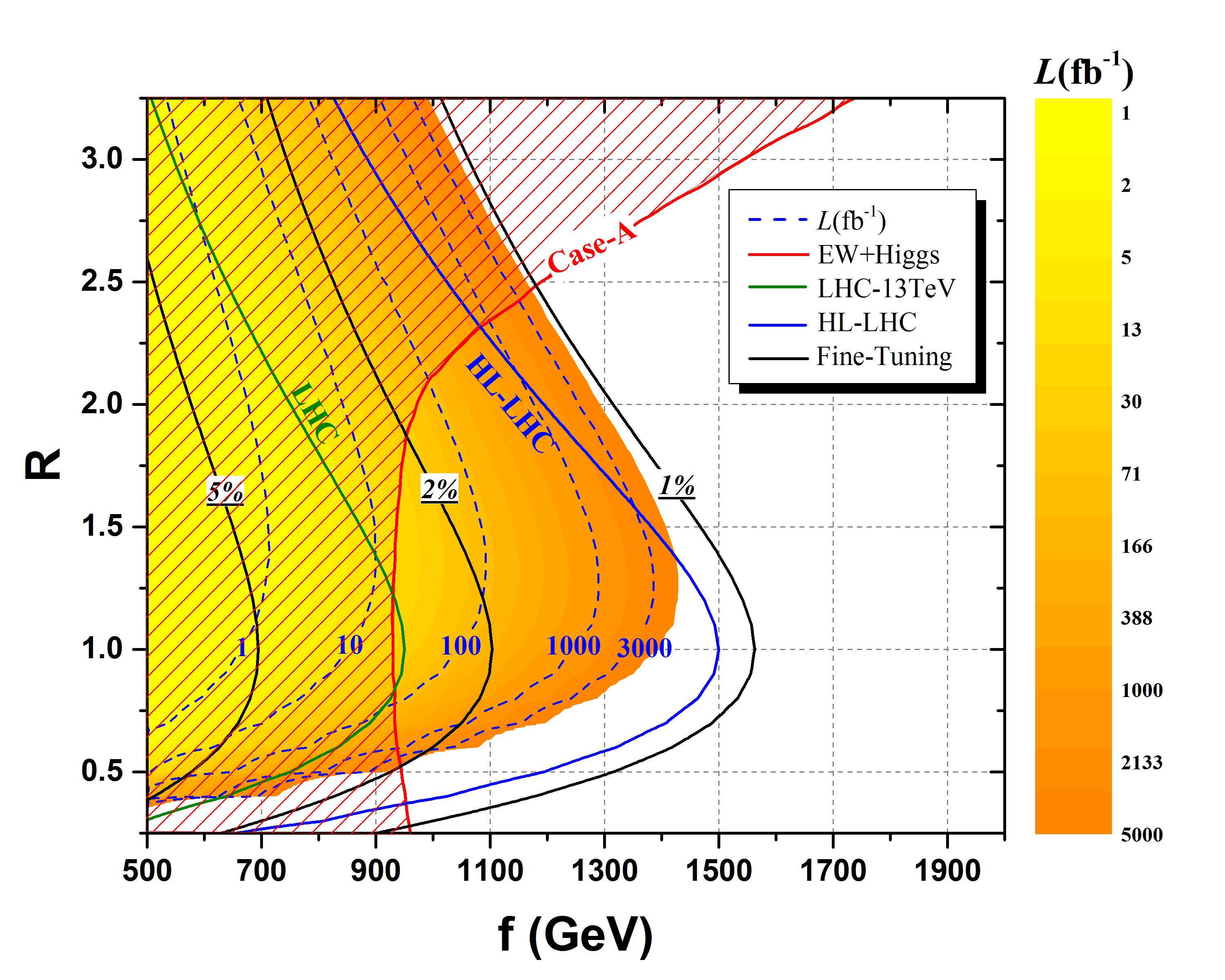}
	\includegraphics[width=0.48\textwidth]{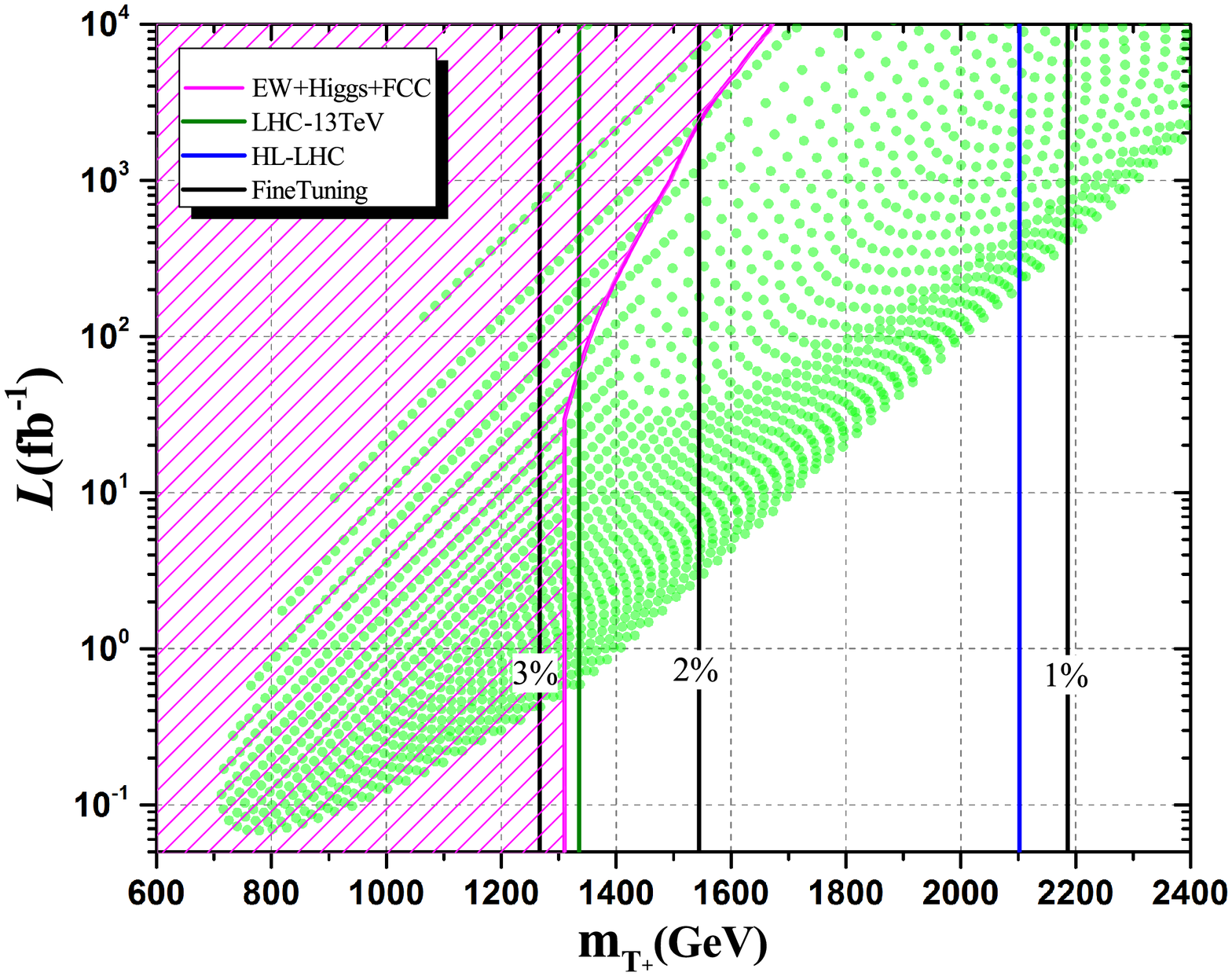}
	\caption{(left) The 2$\sigma$ limit from FCC-eh with polarized $ e^- $ beams and the 2$ \sigma $ limit from the EWPO and Higgs data, the LHC experiment and the fine-tuning in the $ R-f $ plane; (right) The corresponding 2$\sigma$ limits in the $L-m_{T_{+}}$ plane.}
	\label{limit} 
\end{figure}

In Fig.\ref{limit}, we show the 2$\sigma$ limit from FCC-eh with polarized $ e^- $ beams, where the limits from the indirect measurements and the LHC experiments have also been displayed. In order to examine the fine-tuning, we also show the  contours of required fine-tuning 1\%, 2\% and 5\% in Fig.\ref{limit}. To be clear, we show these limits in the $ R-f $ plane(left) and $L-m_{T_{+}}$ plane(right) in the two different panels of Fig.\ref{limit}, respectively. From the left panel of Fig.\ref{limit}, we can see that there will be a considerable $ R-f $ region beyond the current LHC and indirect experiments can be excluded by the FCC-eh with the integrated luminosity $L> 100$fb$^{-1}$. \add{Besides, for the same integrated luminosity of 3000 fb$^{-1}$, we can see that the limit ability of the FCC-eh is mildly weaker than the HL-LHC for $R<1.5$, but better than the HL-LHC for $R>1.5$.} 

In the right panel of Fig.\ref{limit}, we can see that the FCC-eh can exclude the top partner mass $ m_{T_+} $ up to 1350 GeV, 1500 GeV and 1565 GeV with integrated luminosities of 100 fb$^{-1}$, 1000 fb$^{-1}$ and 3000 fb$^{-1}$ at the 2$\sigma$ level based on the limit of the EWPO and the Higgs data. Considering the limits from the FCC-eh with 3000 fb$^{-1}$, which corresponds to the top partner mass $ m_{T_+}>1565$ GeV, we can see that the allowed fine-tuning will reach 2\%. \add{If further considering the HL-LHC limit, the fine-tuning above 1\% will still be allowed.} \add{However, the limit of the HL-LHC shown here is just a rough estimate, we will need full data from the HL-LHC to decide whether naturalness is actually an issue or not. As for the HE-LHC or the FCC-hh, we hope they will shed light on the exploration of new physics. So far, they are still incomplete pre-study schemes and more motivations on the detection capability are needed.}

\section{Summary}

In the LHT model, we investigate the single production of vector-like top partner through $ e^-\ p \rightarrow \nu_e\ \bar{T}_+ (\rightarrow \bar{b}\ W^-)\rightarrow \nu_e (\bar{b}\ \nu_l\ l^- ) \rightarrow l^- + \bar{b} + \slashed E_T$ at the future $ep$ colliders. We calculate the production cross sections with (un)polarized electron beams at the LHeC($ \sqrt{s} $ = 1.98 TeV) and FCC-eh($ \sqrt{s} $ = 5.29 TeV), respectively. In order to study the observability of this signal at the $ ep $ colliders, we perform a fast detector simulation and choose some  kinematic cuts to improve the statistical significance. Besides, we find that the polarized beams can also enhance the statistical significance. For the LHeC collider, the limit on the top partner mass from the search for the $T_{+}$ in this $Wb$ channel is weaker than the current limits from the indirect measurements and the LHC direct searches. For the FCC-eh with polarized $ e^- $ beams, we find that the top partner mass can be excluded up to 1350 GeV, 1500 GeV and 1565 GeV with integrated luminosities of 100 fb$^{-1}$, 1000 fb$^{-1}$ and 3000 fb$^{-1}$ at the 2$\sigma$ level, which is better than the current direct and indirect searches. \add{With the same integrated luminosity of 3000 fb$^{-1}$, we can see that FCC-eh and HL-LHC have different advantages in different parameter spaces, respectively. Although the allowed fine-tuning will drop to less than 2\%, it is still acceptable.}

\section*{Acknowledgement}
We thank Dr. Ruibo Li from Zhejiang University for his enthusiastic discussion and help. This work is supported by National Natural Science Foundation of China (NNSFC) (11405047, 11305049) and the Startup Foundation for Doctors of Henan Normal University (qd15207).


\end{document}